\documentclass[letterpaper,aps,prx,twocolumn,floatfix,amsfonts,amssymb,notitlepage,superscriptaddress,longbibliography]{revtex4-2}

\pdfpageattr {/Group << /S /Transparency /I true /CS /DeviceRGB>>}

\usepackage{amsmath}
\usepackage{amsthm}
\usepackage{amssymb}
\usepackage{graphicx}
\usepackage{tabularx}
\usepackage{color}
\usepackage{xcolor}
\usepackage[normalem]{ulem}
\usepackage[colorlinks=true, linkcolor=blue, citecolor=blue, urlcolor=blue]{hyperref}

\newcommand{\be}{\begin{equation}}
\newcommand{\ee}{\end{equation}}
\newcommand{\bea}{\begin{eqnarray}}
\newcommand{\eea}{\end{eqnarray}}

\def\bs#1\es{\begin{split}#1\end{split}}	\def\bal#1\eal{\begin{align}#1\end{align}}

\begin{document}

\title{Arrow of Time as an indicator of Measurement-Induced Phase Transitions}

\author{Nitay Hurvitz}
\affiliation{Raymond and Beverly Sackler School of Physics and Astronomy, Tel Aviv University, Tel Aviv 69978, Israel}

\author{Alon Kochol}
\affiliation{Raymond and Beverly Sackler School of Physics and Astronomy, Tel Aviv University, Tel Aviv 69978, Israel}

\author{Victor Fleurov}
\affiliation{Raymond and Beverly Sackler School of Physics and Astronomy, Tel Aviv University, Tel Aviv 69978, Israel}
\author{Eran Sela}
\affiliation{Raymond and Beverly Sackler School of Physics and Astronomy, Tel Aviv University, Tel Aviv 69978, Israel}


\begin{abstract}
\noindent 
Measurement-induced phase transitions (MIPTs) in monitored quantum systems are typically diagnosed using entanglement-based measures. Here, we develop a complementary thermodynamic perspective based on the arrow of time (AoT), which arises from the intrinsic irreversibility of the quantum measurements driving these transitions. We study the AoT—defined as the logarithmic ratio of forward and backward trajectory probabilities—across a family of models exhibiting MIPTs.
We find that, like entanglement entropy, the AoT is a nonlinear functional of the averaged density matrix; however, in contrast to entanglement, it is associated with a local operator. To determine whether the AoT exhibits critical behavior, we formulate and exactly solve a model of a random quantum circuit with non-projective measurements. This allows us to analytically demonstrate that the AoT displays nonanalytic behavior and identify its critical exponent.
Our results establish the AoT as a novel diagnostic for phase transitions in monitored quantum systems.
\end{abstract}
\maketitle

\section{Introduction}
Measurement-induced phase transitions (MIPTs) occur in monitored quantum many-body systems due to the competition between unitary dynamics and measurements~\cite{Skinner_2019,Li_2018,Li_2019,Chan_2019,Bao_2020,Jian_2020,Zabalo_2020,Gullans_2020}. They mark a change in the scaling properties of entanglement from a volume law at weak monitoring to an area law at strong monitoring, and are therefore inherently related to quantum information scrambling. They can also be understood from the perspective of the quantum error-correction threshold~\cite{PhysRevLett.125.030505,Li_2021}. 

A natural description of MIPTs is provided by the quantum trajectory framework, in which the system’s evolution is conditioned on sequences of measurement outcomes~\cite{Daley_2014,PhysRevLett.70.2273,wiseman2009quantum}. While the density matrix obtained by averaging over trajectories remains smooth across the transition, individual trajectories exhibit highly nontrivial behavior, including a measurement-rate-driven transition in their entanglement scaling law. This reinforces the view that MIPTs lack a conventional thermodynamic order parameter.

Independently, stochastic thermodynamics has emphasized that quantum measurements are intrinsically irreversible~\cite{Crooks_1999,Seifert_2012,Horowitz_2013,Leggio_2013,Breuer_2003}, and therefore can be quantified on the single trajectory level by the entropy production~\cite{crooks2008quantum,elouard2017role,Dressel_2017,Manikandan_2019,Manikandan_2019_FT,Manzano_2022} or  ``arrow of time" (AoT)~\cite{Dressel_2017}. The AoT is a general measure in any stochastic system, aimed to quantify the irreversibility of random processes. It is defined as the logarithmic ratio between a trajectory's probability and its probability in the time-reversed dynamics, meaning that it is positive 
(“points towards the future”) for any process that is unlikely to be undone, and it turns negative (“reverses”) only rarely, if unlikely events occur. Such trajectory-level measures have been extensively studied in continuously monitored systems 
including experimentally with a single superconducting qubit~\cite{Harrington_2019} or weakly interacting atoms~\cite{jayaseelan2021quantum}.

Despite the central role of measurements in both contexts, the AoT has not been explored 
in many body systems, while MIPTs have been characterized almost exclusively using entanglement-based diagnostics. This raises a natural question: does the arrow of time itself exhibit critical behavior  in many-body systems at the boundary between unitary-dominated and measurement-dominated dynamics? In other words, are the dynamics in the measurement-dominated phase more irreversible than in the unitary-dominated phase?

In this work, we provide a thorough analysis of the arrow of time in model systems undergoing MIPTs.
We begin with a proof of concept demonstrating a transition in the AoT at the level of a single quantum trajectory. We proceed to the continuous measurement limit, which is the primary setting in which the AoT has previously been studied at the level of a single qubit. We generalize these results to the many-body case, showing that, in general, the AoT is associated with a local observable and depends nonlinearly on the system’s density matrix. Then, as our main result, we rigorously show that the AoT undergoes a transition in generic random quantum circuits, and we identify its critical exponents. 

In the context of random quantum circuits, where exact results have enabled the identification of the universality class of MIPTs in a solvable limit with percolation, the computation of the AoT requires a key generalization. This generalization concerns the type of measurements: in the context of the AoT it is essential to avoid projective measurements. 
Projective measurements are fully irreversible and therefore cannot be smoothly characterized by a finite AoT within the microscopic time-reversal framework that we adopt~\cite{Dressel_2017,Manikandan_2019}.  

Replacing projective measurements with invertible measurement operators has, to the best of our knowledge, not yet been achieved within an exact analytic framework capable of identifying critical exponents, as in the projective case~\cite{Skinner_2019,Li_2018,Li_2019,Chan_2019,Bao_2020,Jian_2020,Zabalo_2020,Gullans_2020}. Nevertheless, the nature of the transition is expected to remain unchanged. This expectation is supported by numerical studies~\cite{Li_2019}, as well as by analyses in the limit of continuous measurements~\cite{Jacobs_2006,Szyniszewski_2019}, which show that the transition remains in the same universality class—particularly in chaotic circuits with Haar-random unitary gates, where a transition between volume- and area-law entangled phases is observed. Whether the thermodynamic arrow of time can serve as an alternative indicator of this transition is, however, a subtle question, and establishing its nonanalytic behavior beyond doubt requires exact methods, as achieved here. 

Another case which we touch upon corresponds to one-dimensional (1D) monitored free fermions described by the stochastic Schrodinger equation. As in the extended literature that we now briefly review, precisely identifying the MIPT in these models is subtle. 
Early works on monitored 1D free fermions showed that the volume law phase is unstable to arbitrarily weak measurements~\cite{cao2019entanglement,fidkowski2021dynamical}, and the entanglement scaling crosses over from logarithmic—characteristic of critical 1D systems—to area-law scaling~\cite{alberton2021entanglement,Turkeshi_2021,Piccitto_2022,Szyniszewski_2023}.
It was initially proposed that the transition is of Berezinskii–Kosterlitz–Thouless type~\cite{alberton2021entanglement,buchhold2021effective}, however it was questioned whether this transition survives in the thermodynamic limit~\cite{Coppola_2022} and in fact subsequent developments formulated a replica field-theory description
~\cite{poboiko2023theory}, revealing a close correspondence with Anderson localization in disordered systems in $d+1$ dimensions. Within this mapping the monitored 1D problem corresponds to weak localization in 2D, which is known not to exhibit a true localization transition but instead features an exponentially large correlation length. 

As we will discuss below, the AoT is directly linked to a type of correlation function which is non-linear in the density matrix that entered the debate on the nature of the MIPT, see Eq.~(\ref{eq:C}) below. Although numerically it was initially believed to exhibit a power law phase for free fermions with a conserved charge, the exact replica limit approach proved that observing the critical phase strictly requires to remove the $U(1)$ symmetry as in Refs.
~\cite{Turkeshi_2021,Piccitto_2022,Poboiko_2024,Fava_2023} (switching 
from 
localization to weak-antilocalization). Alternatively, the critical phase appears for monitored free fermions above 1D~\cite{Poboiko_2024,Chahine_2024,chahine2025spectraltransitionsentanglementhamiltonian} 
- or by adding interactions~\cite{muller2025monitoredinteractingdiracfermions}. As we find, the average AoT is the local limit of this correlation function.  

\subsection{Extended outline}
This paper is organized as follows. In Sec.~\ref{se:definition} we define the AoT in  monitored quantum many body systems. 
To each trajectory we associate an AoT, which quantifies whether a given sequence of outcomes is more likely to occur forward in time or in reverse. For a single qubit under continuous measurement, this notion is well understood: measurement tends to project the state toward an eigenstate of the measured observable, while rare fluctuations effectively “undo” the measurement and drive the system away from it~\cite{Katz_2008}. The ratio between these forward and backward probabilities determines the AoT. Extending this trajectory-resolved AoT to many-body systems is nontrivial, and the presence of a MIPT provides a natural arena in which to investigate it.

In Sec.~\ref{se:no-click} we apply this definition for monitored systems described by the no-click limit, which is special in that it corresponds to a single  quantum trajectory. This allows for a direct and unambiguous application of the microscopic definition of the AoT, without the need to average over exponentially many
trajectories. Already in this setting we uncover several  results: At long times, the AoT maps onto the steady-state expectation value of the monitored measurement operator—a local observable, in sharp contrast to entanglement entropy, which is intrinsically nonlocal. This local quantity, which is directly accessible in single-qubit experiments, exhibits a nonanalyticity both at the single-qubit level and at the MIPT when generalized to many qubits. 

Moreover, the no-click dynamics is governed by an effective non-Hermitian Hamiltonian, generating deterministic evolution with complex eigenvalues. We establish an exact relation between the  AoT production rate and the eigenvalue of this many-body non-Hermitian Hamiltonian with the largest imaginary part—namely, 
the eigenmode that dominates the long-time dynamics. 

In Sec.~\ref{se:cont_meas} we make a general analysis of the AoT including its mean and variance in continuously measured many-body systems. We derive an expression for the many-body AoT  as a stochastic time integral involving the measurement record and the evolving quantum state—a direct generalization of the single-qubit result~\cite{Dressel_2017}. The AoT averaged over the measurement outcomes is proportional to the stochastic average of the square of a quantum expectation value  (of the measured operator $\sigma^z$). This quantity is thus nonlinear in the density matrix and maps directly onto correlators previously studied primarily in the context of MIPTs in free-fermion systems,
\be
\label{eq:C}
C({\bf{x}}-{\bf{x}}')= \overline{ \langle \sigma_{\bf{x}}^z \sigma_{\bf{x}'}^z  \rangle}-\overline{ \langle \sigma_{\bf{x}}^z   \rangle  \langle \sigma_{\bf{x'}}^z   \rangle }.
\ee
However, since the measurements are local, the average AoT probes only the equal-time, equal-position limit of these correlations (${\bf{x}}=(x,t)=(0,0)$). Careful numerical work is needed to test whether this local correlator displays non-analytic behavior.
In contrast, fluctuations of the AoT probe precisely this correlation function at distinct points in space and time.

In Sec.~\ref{se:random_quuantum_circuits} we generalize exact solvability of MIPTs in random quantum circuits with unitary gates drawn from a Haar ensemble~\cite{Skinner_2019,Li_2018,Chan_2019,Bao_2020,Gullans_2020}. A series of works using this framework identified the phase transition between area-law and volume-law entanglement phases. Here we employ these analytical methods and the associated mappings to statistical-mechanics models to investigate the role of the AoT. To this end, we replace projective measurements with invertible ones within the random-circuit framework, in a way that maintains exact solvability. 

We show that the AoT  can be expressed exactly as a partition function upon taking an appropriate replica limit. Following Refs.~\cite{Bao_2020,Jian_2020}, we evaluate this limit for large  local Hilbert-space dimension. We find that as in the projective measurement case, the problem maps explicitly onto a percolation model. Interestingly, the AoT —unlike entanglement, which is a correlation quantity—does not depend on spatial separation and instead behaves as a thermodynamic quantity. In the percolation mapping, it appears to play the role of the average cluster size, with critical exponent of the specific heat $\alpha$, reinforcing the thermodynamic nature of the AoT. 

We summarize in Sec.~\ref{se:summary}. In essence, our key finding is that the AoT gives a glimpse directly into the critical properties of MIPTs from the perspective of the  bulk thermodynamic critical free energy of the theory - rather than the order parameter of the theory~\cite{Gullans_2020micro} which explicitly probes spreading of quantum information. We also discuss how the AoT puts MIPT in a more favorable state with respect to the postselection problem.

\section{Quantum trajectories and arrow of time}
\label{se:definition}
We begin by reviewing the definition of the arrow of time introduced by Dressel et al.~\cite{Dressel_2017} and elaborated on in subsequent works ~\cite{Manikandan_2019,
Manikandan_2019_FT,Yanik_2022}, based on the possibility to apply microscopic time reversal generating a backward quantum trajectory which retraces the same path in Hilbert space as the forward trajectory, but in reverse temporal order. 

We consider a system of $L$ qubits located at positions $x=1,\dots,L$, evolving in discrete time steps labeled by $\tau=1,\dots,N$. Each time step consists of a measurement followed by a specified unitary evolution operator $U$ (which may vary between time steps). Each measurement acts locally and independently on each qubit, and is described by a family of single qubit Kraus operators $\{A_i\}$ satisfying the completeness relation $\sum_i A_i^\dagger A_i = \mathbb{I}$. Conditioned on the measurement outcome $i$, which occurs with probability $p_i = \langle \psi| A_i^\dagger A_i |\psi \rangle $, a state $|\psi\rangle$ transforms via  $|\psi \rangle \to \frac{A_i|\psi \rangle}{\sqrt{p_i}} $.

A quantum trajectory is specified by a  sequence of measurement outcomes $m=\{i_{x,\tau}\}$ on qubit $x$ at time step $\tau$. The probability of each quantum trajectory depends on the initial state $|\psi_0 \rangle$ and is given by $p_m={\rm{Tr}}\rho_0 A^\dagger_m A_m$ where the total evolution operator including measurements and unitary dynamics is the time ordered product $A_m =\mathcal{T}\left( \prod_{\tau=1}^{N}U_\tau \times \prod_{\tau=1}^{N}  \otimes_{x=1}^L A_{i_{x,\tau}}    \right)$ satisfying $\sum_m A^\dagger_m A_m=\mathbb{I}$. 

To define an arrow of time for a trajectory $m$ consider the backward trajectory which starts at the final state of the forward trajectory $| \psi_f \rangle = A_m | \psi_0 \rangle$. It can be normalized as
\be
|\psi_{f}\rangle = \frac{A_m |\psi_{0}\rangle}{\sqrt{p_m}},
\ee
where $p_m=\langle \psi_0|A^\dagger_m A_m |\psi_0 \rangle$ is the probability of the forward trajectory. Note that $p_m$ is a joint probability of all the measurement outcomes and it can be expressed as a product of probabilities conditioned by the previous measurement outcomes.

It has been demonstrated in ~\cite{Manikandan_2019} that the quantum trajectory
of a qubit can be reversed by applying the Kraus operators given
by
\be
\label{eq:reversed_op}
\tilde{A}_i = \Theta A_i \Theta^{-1},
\ee
on the final state $|\psi_{f}\rangle$, in reversed order. Here $\Theta$ is the time-reversal operator, which in the case of rank-2 Kraus operators ensures $\tilde{A}_j A_j \propto \mathbb{I}$.

Namely acting with the operator $\tilde{A}_m =\tilde{\mathcal{T}}\left( \prod_{\tau=1}^{N} \tilde{U}_\tau \times \prod_{\tau=1}^{N}  \otimes_{x=1}^L \tilde{A}_{i_{x,\tau}}    \right)$ on $|\psi_{f}\rangle$, one retracts the same intermediate states in Hilbert space but in reverse order and eventually recovers $|\psi_0\rangle$ up to normalization. Here $\tilde{\mathcal{T}}$ is the anti-time ordering operator and $\tilde{U} = \Theta U \Theta^{-1}$.

The backward dynamics is also stochastic, so the exact trajectory is traced back only with certain probability
\be
\tilde{p}_m=\langle \psi_f |\tilde{A}_m^\dagger \tilde{A}_m|\psi_{f}\rangle.
\ee
The arrow of time associated with a quantum trajectory is then defined as the ratio of those forward-in-time and backward-in-time probabilities,
\be
Q_m = \log \frac{p_m}{\tilde{p}_m}.
\ee
It is thus crucial for this definition of the AoT that the trajectory would be reversible in Hilbert space with non-zero probability. For this reason, the measurement operators are required to be invertible, i.e not projective. One can see for example the issue with a projective measurement of a qubit in the state $|+\rangle = \frac{1}{2}(|0\rangle+|1  \rangle)$ to either $|0\rangle$ or $|1\rangle$ - none of the time-reversed measurement operators (Eq. \ref{eq:reversed_op}) will restore the superposition, thus $\tilde{p}_m=0$ and the AoT diverges. However, as long as the measurement operators are invertible, even if close to the projective limit, the 
backward-in-time trajectory exists with non-zero probability and the AoT remains finite.

The average arrow of time is then 
\be
\label{eq:av_Q}
\overline{ Q}=\sum_m p_m \log \frac{p_m}{\tilde{p}_m}.
\ee
We note that while $\sum_m p_m=1$, the same is not true for $\tilde{p}_m$ because for every forward trajectory $m$, $\tilde{p}_m$ is the probability of returning to the initial state precisely via the reversed path, out of many other paths that start in the final state $|\psi_f\rangle$ which changes according to $m$. The probability $\tilde{p}_m$ is thus part of a different distribution for every $m$ in the summation. Therefore, the average AoT in Eq.~(\ref{eq:av_Q}) is not a relative entropy. 

We also note that the definition of the AoT based on reversing the path in Hilbert space should be distinguished from a similar definition of entropy production~\cite{crooks2008quantum,Manzano_2015,elouard2017role} which constructs the backward process by reversing only the sequence of measurement outcomes, without requiring that the quantum state follows the time-reversed evolution.

\section{No-click dynamics
}
\label{se:no-click}
In this section, we study the
an arrow of time in measurement-induced phase transitions by focusing on the no-click limit of monitored dynamics~\cite{Turkeshi_2021,Malakar_2024,zerba2023measurementphasetransitionsnoclick,paviglianiti2023multipartiteentanglementmeasurementinducedphase}. The no-click trajectory can indeed be identified experimentally. For example using a single superconducting qubit, the experiments in~\cite{Minev_2019, guttel2026graduallyopeningschrodingersbox} demonstrate individual realizations of stochastic evolution — which display discrete jumps (click/emissions events), and continuous evolution between the jumps, conditioned on no detection. Despite the absence of detection events, the system evolves under a measurement-conditioned dynamics: the evolution is continuously monitored and conditioned on no clicks occurring. This dynamics is non-unitary and generated by an effective non-Hermitian Hamiltonian, making it intrinsically irreversible.

\subsection{No-click dynamics of a single qubit}
To introduce the measurement model, we first consider a single qubit and then generalize to many qubits undergoing a MIPT.  For a single qubit, we consider the Kraus operators
\be
A_0 = \begin{pmatrix}
1 & 0 \\
0 & \cos \epsilon
\end{pmatrix}, \quad
A_1 = \begin{pmatrix}
0 & 0 \\
0 & \sin \epsilon
\end{pmatrix},
\ee
which model a detector that couples only to the qubit's excited state, such that when the qubit is excited the detector may click with a small probability $\sin^2\epsilon$. Outcome $i=1$ (click) indicates that the detector registered the excited state $|e\rangle = |\downarrow \rangle= (0,1)^T$. Since clicks can never be produced in the ground state, this corresponds to a projection on $|e\rangle$. Outcome $i=0$ (no-click) indicates that the detector did not register the excited state.  Since the absence of clicks is possible in both states, this only slightly increases the probability to be in the ground state  $|g\rangle= |\uparrow \rangle = (1,0)^T$.

The no-click dynamics corresponds to the post-selected quantum trajectory in which  $A_0$ exclusively acts at every time step. For small $\epsilon$, we use a standard definition of a measurement rate $\gamma$ and an infinitesimal time step $dt$ via $\epsilon^2 = \gamma\, dt$, so that a no-click trajectory corresponds to evolution under a non-Hermitian Hamiltonian,
\be
\label{eq:no_click_eff}
A_0 = I - \frac{\gamma dt}{2} |e\rangle \langle e|= e^{-\frac{\gamma}{2}\hat{n} dt},  
\ee
where we denoted the measurement operator $\hat{n} = |e \rangle \langle e|=(1-\sigma^z)/2$.
This operator suppresses the excited state amplitude and increases the AoT, as we will see below.  Let us also define the excited state population 
$n={\rm{Tr}} \rho \hat{n}$,  and $z={\rm{Tr}} \rho \sigma^z=1-2n$.

{\bf{Arrow of time and local expectation value:}} Considering a single time step $dt$ of the effective Hamiltonian, starting from a wave function $| \psi_\tau \rangle$ at time $\tau$, the probability of the no-click event 
is
\be
p_0=\langle \psi_\tau| A_0^\dagger A_0 | \psi_\tau \rangle =1-\gamma n dt.
\ee
The probability of the backward trajectory is defined using the time-reversed measurement operators~\cite{Manikandan_2019}:
\be
\tilde{A}_0 = \begin{pmatrix}
\cos \epsilon & 0 \\
0 & 1
\end{pmatrix},~~~\tilde{A}_1 = \begin{pmatrix}
\sin \epsilon & 0 \\
0 & 0
\end{pmatrix}.
\ee
While a click is projective and thus not invertible ($\tilde{A}_1 A_1 = 0$), the no-click evolution is invertible with probability
\be
\tilde{p}_0= \langle \psi_{\tau+1}|\tilde{A}^\dagger_0 \tilde{A}_0|\psi_{\tau+1} \rangle,
\ee
where $|\psi_{\tau+1} \rangle=\frac{A_0 | \psi_\tau \rangle}{\sqrt{p_0}}$.
It then follows that the increment of the AoT, $dQ = \frac{p_0}{\tilde{p}_0}$, is given by $\frac{dQ }{dt}=\gamma (1 - 2 n) = \gamma z$, leading to the arrow of time of
\be
\label{eq:rho_ee} 
Q(t) = \gamma t (1 - 2 \langle n \rangle_t) = \gamma t \langle z \rangle_t,
\ee
where $\langle \cdot\rangle_t = \frac{1}{t}\int_0^t\cdot~dt'$ is an average over time. We see that under microscopic time reversal, the AoT depends on the excited state population. It is increasing (decreasing) for $n < 1/2$ ($n > 1/2$). This result gives the AoT per qubit independently of the Hamiltonian that controls the unitary dynamics, which simply determines the excited state population $n$. If the occupation of the excited state increases, a click becomes more likely and the probability of a no-click trajectory decreases. We emphasize that here the AoT refers to a single trajectory, not an average over all trajectories, as will be explored in later sections.

\begin{figure}
\centering
\includegraphics[width=\columnwidth]{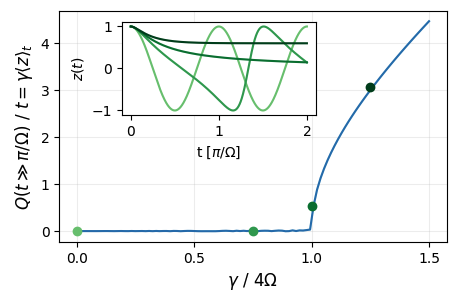}
\caption{Example of a transition in the AoT for a single qubit evolving under the Hamiltonian in Eq.~(\ref{eq:H_1qubit_nonhermitian}). Plotted is the AoT at long times divided by time
. Inset: time evolution of the excited-state probability $n(t)$, showing saturation at $z>0$ above a critical measurement rate.
}
\label{fig:fig_no_click_Zeno}
\end{figure}

{\bf{Example - Zeno transition:}} To illustrate the AoT intuitively, we consider the standard Zeno transition in a single qubit under a Rabi coherent drive, together with the no-click monitoring dynamics. The overall evolution is then given by the non-Hermitian Hamiltonian
\be
\label{eq:H_1qubit_nonhermitian}
H= \Omega \sigma^x -i \frac{\gamma}{2} \hat{n}.
\ee
The time dependence of the excited-state probability is shown in the inset of Fig.~\ref{fig:fig_no_click_Zeno}, highlighting a transition from a Rabi phase, where the probability oscillates between 0 and 1, to a Zeno phase for $\gamma \ge 4\Omega$, where at long times the qubit saturates to a fixed excited-state probability. Equation~(\ref{eq:rho_ee}) immediately implies that the long-time AoT vanishes in the Rabi phase, but is finite in the Zeno phase, as seen in Fig.~\ref{fig:fig_no_click_Zeno}, exactly quantifying the time-symmetric nature of the Rabi oscillations, compared to the irreversibility related with reaching the Zeno state.

This simple dynamical transition~\cite{Snizhko_2020} therefore has an immediate signature in the arrow of time and can be tested in experiment~\cite{guttel2026graduallyopeningschrodingersbox}.

\subsection{
Ising chain under no-click monitoring}

We now generalize to $L$ spins interacting via an Ising Hamiltonian, with the same local measurement model applied to each site. In the no-click limit, the dynamics are governed by the effective non-Hermitian Hamiltonian
\be
\label{eq:H_eff}
H_{\mathrm{eff}} = J \sum_{j=1}^{L} \sigma^x_j \sigma^x_{j+1} - \frac{i \gamma}{2} \hat{n}_j,
\ee
where we use periodic boundary conditions. This model exhibits a type of MIPT at $\gamma_c = 4J$ and $L\to\infty$~\cite{Turkeshi_2021}, which can be understood as a many-body Zeno transition ~\cite{Biella_2021}. Most importantly, it is exactly solvable via a Jordan–Wigner transformation~\cite{Lee_2014}. Since the dynamics are non-unitary, all eigen-states are in general decaying, and so the long-time behavior of this model is simply characterized by the longest-living eigen-state. The Zeno transition could then be viewed as the convergence of this eigen-state for $\gamma >\gamma_c$ into a disentangled eigen-state of the measurement operators, i.e. the Zeno state. 
Here, we show that this transition could also be seen as a non-analytic behavior of the AoT, being the first evidence for the AoT as an indicator of MIPTs.

\begin{figure}
\centering
\includegraphics[width=\columnwidth]{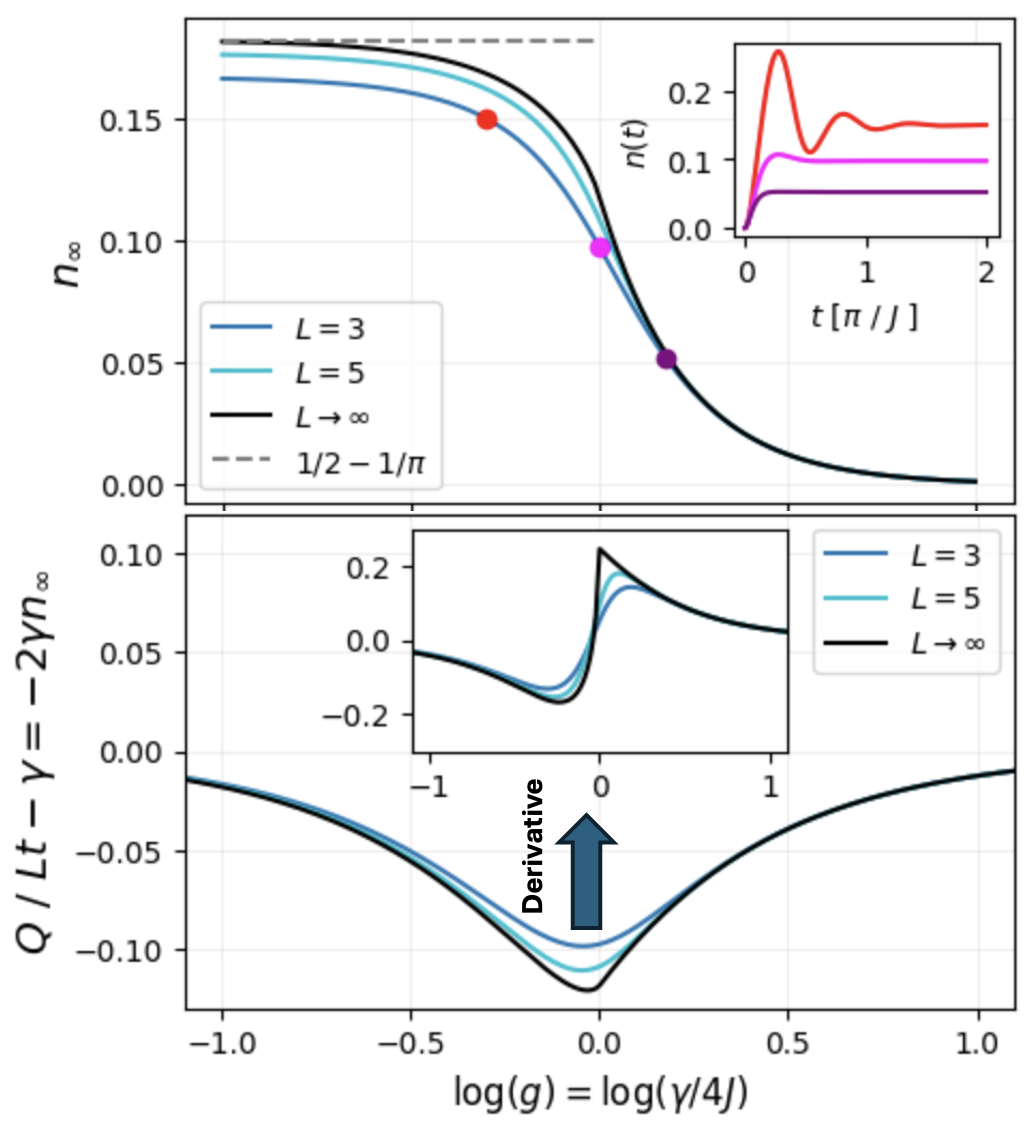}
\caption{(top) Long-time occupation of the excited state for the no-click trajectory in Ising dynamics, starting from the 
ground state. 
(bottom) Arrow of time per spin divided by time as obtained from Eq.~(\ref{eq:no_click_aot}). For clarity, we subtract the single-spin contribution $\gamma$ to isolate the many-body part. This quantity exhibits a sharp change in slope at $g_c=1$. A nonanalytic behavior becomes visible in its derivative for sufficiently large system sizes. 
}
\label{fig:fig_tent}
\end{figure}

The simple generalization of the AoT from Eq.~(\ref{eq:rho_ee}) to $L$ spins reads 
\be
\label{eq:no_click_aot}
Q(t) = \gamma t \sum_{x=1}^L (1 - 2 \langle n_x \rangle_t)  \xrightarrow{n_x = n}  \gamma L t (1 - 2 \langle n \rangle_t),
\ee
where the last relation is obtained in the translationally invariant case, $n_x = n$, as obtained for Eq.~(\ref{eq:H_eff}) with periodic boundary conditions (the present model does not undergo spontaneous symmetry breaking). Furthermore, at long times the excited state occupation $n$ reaches a steady state value $n_\infty$ (see inset of Fig.~\ref{fig:fig_tent}(a)) so we can replace $\langle n \rangle_t \ \to n_\infty$.
Combining this with a spectral analysis of $H_{\mathrm{eff}}$ shown in Appendix~\ref{appendix:non-hermitian}, see Eq.~(\ref{eq:Sigma_Gamma}), we arrive at a relation between $n_\infty$ and the imaginary part of the longest-living eigenvalue of the non-Hermitian many-body Hamiltonian,
\be
\label{eq:rel}
\frac{\Gamma_{\min}}{L} = \frac{\gamma n_\infty}{2}.
\ee
This relation, previously pointed out for a single qubit~\cite{Lee_2014}, is exactly confirmed for our model using the Jordan-Wigner transformation. The dependence of the long-time expectation value $n_{\infty}$ on the measurement strength, parameterized by $g = \frac{\gamma}{4J}$ (with the MIPT occurring at $g_c = 1$~\cite{Turkeshi_2021}), is shown in Fig.~\ref{fig:fig_tent} (top) and for $L \to \infty$ is given by
\be
n_{\infty} = \int_0^\pi \frac{dk}{\pi} \frac{\sin^2 k}{\sin^2 k + | \cos k - i g + \sqrt{1 - g^2 - 2 i g \cos k} |^2}.
\ee
It satisfies $n_{\infty} \to \frac{1}{2} - \frac{1}{\pi} \approx 0.1817$ as $g \to 0$, and $n_{\infty} \sim \frac{1}{8 g^2}$ as $g \to \infty$. We note that $n_{\infty}(g \to 0) \ne n_{\infty}(g=0)$. In the limit $g \to 0$ the oscillations toward $n_{\infty}$ persist indefinitely, making the long time effect of $g$ nonperturbative. Below the critical measurement rate, $n_{\infty}$ exhibits a nonanalytic behavior, $n_{\infty}(1+\epsilon) = 0.119 + \mathcal{O}((-\epsilon)^{3/2})$ for $\epsilon < 0$.

In Fig.~\ref{fig:fig_tent} (bottom) we plot the AoT per spin and time. 
We subtract   the trivial single-spin contribution $\gamma L$ to obtain the term proportional to $\Gamma_{{\rm{min}}}$ (a quantity also considered in Ref.~\cite{Biella_2021}). 
The result forms an inverted tent shape, already visible for small system sizes (e.g., $L=5$ in the figure). In the weak-measurement phase, where entanglement was surprisingly shown to be increasing with a finite measurement strength \cite{Biella_2021}, we show that the AoT, in a similar way, is  suppressed by the measurements, up until the critical point from which a steady increase is recovered. The nonanalyticity at $g=1$, which becomes apparent upon taking a derivative with respect to $g$, sharpens with increasing system size.

In summary, the arrow of time in the no-click Ising model remains finite for all values of $g$, reflecting the persistent presence of measurement on both sides of the transition. Nevertheless, it exhibits a clear nonanalyticity at the critical point. Therefore, the entanglement transition previously identified in this model—from logarithmic to area-law scaling at $g_c = 1$—is accompanied by a subtle 
nonanalytic signature in the arrow of time, a thermodynamic quantity. Unlike entanglement, however, this quantity is governed by a local expectation value.

\section{Continuous Measurements}
\label{se:cont_meas}
In this section, we consider systems evolving under continuous measurements. In this case, instead of a single no-click trajectory, there will be infinitely many possible trajectories, each with a different value of the AoT. So the AoT will have to be analyzed statistically.
A convenient platform to analyze the AoT for a single qubit under continuous measurements is superconducting qubit experiments, for a recent book see Ref.~\cite{jordan2024quantum}. Here the qubit is dispersively coupled to a microwave  cavity, and the qubit state is encoded in the phase of a microwave tone that probes the cavity, giving a continuous measurement record to reconstruct the individual quantum trajectories of the qubit state~\cite{Weber_2016}. In this single qubit context the AoT had been considered both theoretically~\cite{Dressel_2017} and experimentally~\cite{Harrington_2019}. 
We provide simple generalizations from the single-qubit case. These allow us to relate the AoT—including its mean value and variance—to correlation functions of the form of Eq.~(\ref{eq:C}) of the local measured observable, which are non-linear in the system’s density matrix.

We start from a discrete measurement model of a qubit with a pair of Kraus operators
\be
\label{eq:Apm}
A_{+1} = \frac{1}{\sqrt{1 + a^2}}
\begin{pmatrix}
a & 0 \\
0 & 1
\end{pmatrix},
\qquad
A_{-1} = \frac{1}{\sqrt{1 + a^2}}
\begin{pmatrix}
1 & 0 \\
0 & a
\end{pmatrix},
\ee
where $a = 1 + \epsilon$. $A_{j=\pm 1}$ gradually projects the  qubit towards its $\sigma^z = \pm 1$ eigenvalue. Each measurement outcome $j= \pm 1$ occurs with a probability that depends on the pre-measurement state \be
\label{eq:ps}
p_{\pm 1} = \langle \psi_\tau | A_{\pm 1}^\dagger A_{\pm 1} | \psi_\tau \rangle = \frac{1}{2}(1 \pm \epsilon z_\tau) + \mathcal{O}(\epsilon^2),
\ee
where $z_\tau=\langle \psi_\tau|\sigma^z|\psi_\tau \rangle$.  
Continuous measurement corresponds to the $\epsilon \to 0$ limit with a large number of steps $N=T/dt \to \infty$ such that $\epsilon^2 N/T$ remains constant. We define the measurement rate $\gamma$ via $\epsilon^2=\gamma dt \ll 1$. 

Focusing initially on a single qubit, consider a measurement trajectory $m=\{ j_\tau \}$ ($\tau=1,\dots,N$) in the limit of $\epsilon, dt \to0$. 
The  probability of the forward trajectory is
\be
p_m=\left( \frac{1}{2} \right)^N \prod_{\tau=1}^N (1+\epsilon j_\tau z_\tau).
\ee
The Kraus operators in Eq.~(\ref{eq:Apm}) satisfy $\tilde{A}_{+1} = A_{-1}$ and $\tilde{A}_{-1} = A_{+1}$. Therefore the backward path probability is
\be
\tilde{p}_m=\left( \frac{1}{2} \right)^N \prod_{\tau=N}^1 (1-\epsilon j_\tau z_{\tau+1}).
\ee
The ratio of these probabilities gives the arrow of time,
\be
\label{eqLQ_m_j_tau}
Q_m=\log \frac{p_m}{\tilde{p}_m} =  2\epsilon \sum_{\tau=1}^N j_\tau z_\tau^{\rm{Strat}},
\ee 
where  $z_\tau^{\rm{Strat}}=\frac{z_\tau+z_{\tau+1}}{2}$. Notice the similarity to Eq.~(\ref{eq:rho_ee}) which corresponds to a single trajectory with $2\epsilon j_\tau \to \gamma$, hinting at a general structure for any type of continuous monitoring.

The AoT in Eq.~(\ref{eqLQ_m_j_tau}) depends on the random measurement outcomes $j_\tau=\pm$. It was shown by Dressel et al.~\cite{Dressel_2017} that the 
trajectory average AoT, $\overline{Q}$, can be expressed solely in terms of the trajectory average of $z_{\tau}^2$. Since the derivation, based on stochastic calculus, remains intact for the many qubit case, we directly state the result,
\bea
\label{eq:mean_z2}
\overline{Q}  = \epsilon^2 \sum_{x,\tau}  \left(  \overline{z_{x,\tau}^2}  + 1\right),
\eea
see Appendix~\ref{appendix:mean} for details. In the  continuous time limit this gives~\cite{Dressel_2017} $\overline{Q}=\gamma \sum_x\int_0^T dt \left( \overline{z_{x}(t)^2} + 1\right)$. 

This equation should be supplemented with the stochastic Schrodinger equation (SSE) that governs the dynamics,
\bea
&&|\psi_{\tau+1} \rangle-|\psi_{\tau} \rangle=-idt H | \psi_\tau \rangle +       \\&&\sum_{x=1}^L\left[\xi_{x,\tau }\epsilon \left( \frac{\sigma^z_{x,\tau }-z_{x,\tau }}{2} \right) - \frac{1}{2} \epsilon^2 \left( \frac{\sigma^z_{x,\tau }-z_{x,\tau }}{2} \right)^2 \right]| \psi_\tau \rangle,\nonumber
\eea
where $\overline{ \xi_\tau } =0$, $\overline{ \xi_\tau \xi_{\tau'} }=\delta_{\tau \tau'}$ is a white noise related to $j_\tau$ by a shift making $\xi_\tau$ unbiased.

\subsection{Relation to connected correlation function}
As identified in the original studies of MIPTs, a basic object that reflects  MIPTs is the connected correlation function in Eq.~(\ref{eq:C})
~\cite{alberton2021entanglement,poboiko2023theory,buchhold2021effective, M_ller_2022,Chahine_2024}, 
\be
C({\bf{x}}-{\bf{x}}')= \overline{ \langle \sigma_{\bf{x}}^z \sigma_{\bf{x}'}^z  \rangle}-\overline{ \langle \sigma_{\bf{x}}^z   \rangle  \langle \sigma_{\bf{x'}}^z   \rangle }.
\ee
Here it is assumed that a steady state was achieved.
The second term is nonlinear in the density matrix. It is crucial to include this term because  correlations 
which are linear in the density matrix are trivial at long times  where the monitored dynamics leads to the infinite temperature 
state. 

The case ${\bf{x}}={\bf{x}}'$, corresponding to the autocorrelation function, determines the average AoT via Eq.~(\ref{eq:mean_z2}), 
\be
\overline{Q}  = \gamma L T   \left(  2-C(0)\right),~~~C(0)= 1-\overline{ z_{x,\tau}^2}.
\ee
What is known about this correlator? Alberton et al.~\cite{alberton2021entanglement} found numerically that it transitions from an exponential decay in the measured phase to a conformally invariant $1/|{\bf{x}}-{\bf{x}}'|^2$ behavior in the entangling critical phase. This was further elaborated using bosonization~\cite{buchhold2021effective}. While these two behaviors correspond to the long-wavelength limit, for the AoT the limit of the autocorrelation function $C(0)$ is required, also looked at by Alberton et al.~\cite{alberton2021entanglement}. Yet it remain unclear whether it exhibits nonanalyticity versus measurement rate.

\subsection{Numerical results}
\begin{figure}
\centering
\includegraphics[width=\columnwidth]{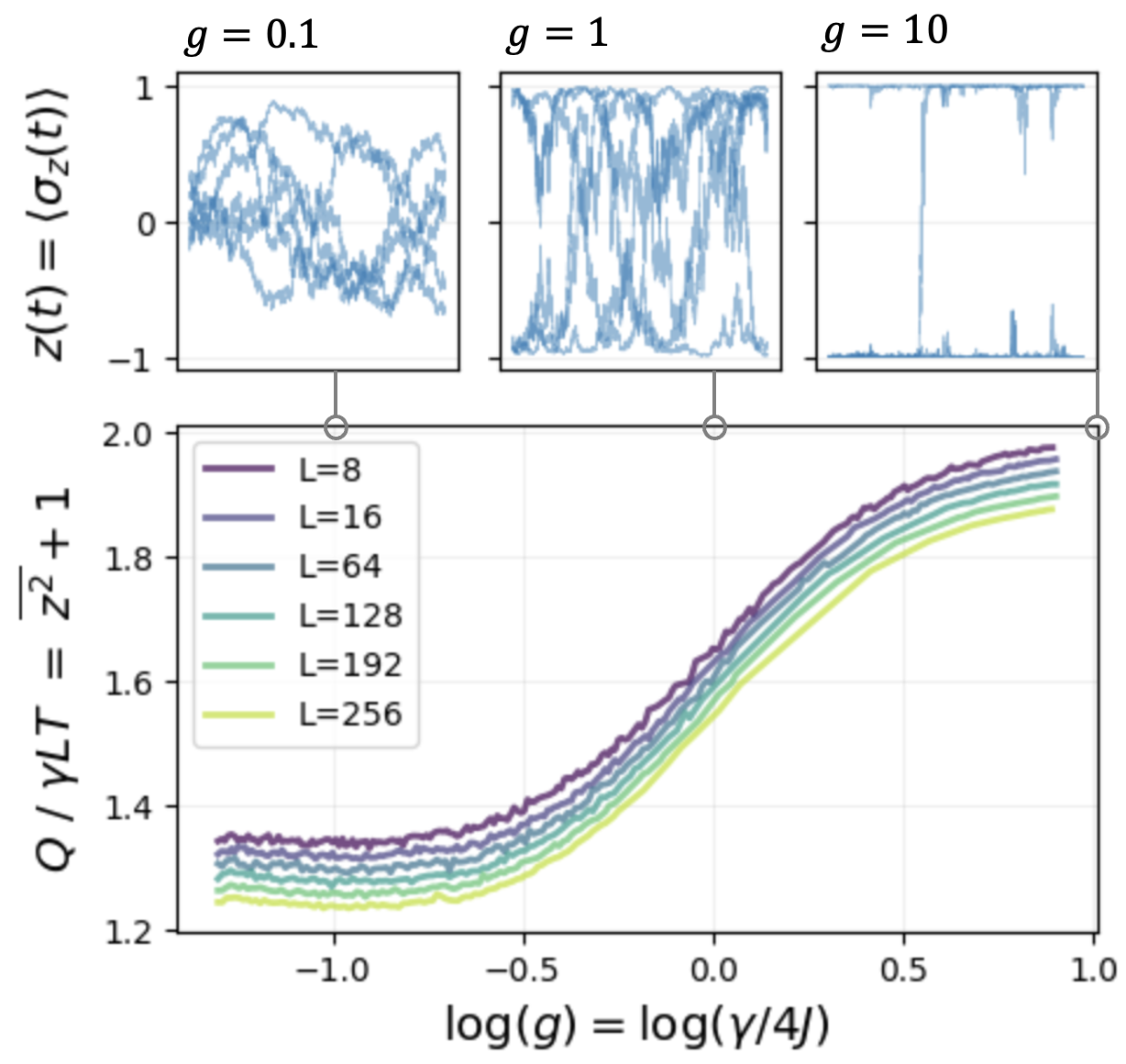}
\caption{Average arrow of time $Q$ divided by measurement rate $\gamma$ and total time $T$ for 
$L$ qubits evolving according to the stochastic Schrodinger equation
with the Ising Hamiltonian. (top) 3 example trajectories for $L=8$ at $g = \gamma/4J \in \{0.1, 1,10\}$ presented by the expectation value of $\hat{\sigma}_z$ of each of the qubits through time. (bottom) Results for $\overline{z^2} + 1$ as a function of $\log g$ with $g=\gamma/4$, for $L\in\{64,128,192,256\}$, each separated by $-0.02$ with respect to $L=8$ for clarity.  
The transition region appears near $\log g =0$, with saturation towards the upper reference level $2$ at large $g$.}
\label{fig:mean_arrow_of_time}
\end{figure}
 In Fig.~\ref{fig:mean_arrow_of_time} we show the average AoT production rate for $L$ qubits evolving according to the SSE with Ising Hamiltonian $H= J\sum_{j=1}^{L} \sigma^x_j \sigma^x_{j+1}$, versus $g=\frac{\gamma}{4J}$. The calculation is done  using Eq.~(\ref{eq:mean_z2}) where quantum trajectories are generated using the correlation function method~\cite{cao2019entanglement}. 
The plotted $y$ axis is simply  $\overline{z^2} + 1 =2-C(0)$, where $\overline{z^2}$ refers to $\langle z_{x,\tau} \rangle^2$ averaged over time $\tau$, space $x$, and over a few trajectories. 
It tends towards $2$ in the Zeno phase, as all qubits get pinned to $z=\pm1$ (see Fig.~\ref{fig:mean_arrow_of_time}, top right).
We can see that the mean AoT displays a crossover across the MIPT previously found in this model~\cite{Turkeshi_2021} and no sharp features can be identified from this data. 

As we have seen in the no-click case, nonanalytic behavior in the AoT may appear in higher-order derivatives with respect to measurement rate. As we will show in the next section, the AoT in generic random circuits exhibits nonanalytic behavior in its second derivative. Moreover, as one can see in Fig.~\ref{fig:n_p} below, when looking directly at the AoT one hardly sees any length dependence, as in Fig.~\ref{fig:mean_arrow_of_time}. Consequently, the present numerical results cannot rule out such nonanalytic behavior. A direct numerical calculation of derivatives of the AoT is essential to assess the presence or absence of nonanalyticity of the AoT in this model; this is left for future work.

Before closing the section we remark that the relation between the AoT and the connected correlation function extends beyond the mean value of the AoT. 
In Appendix~\ref{appendix:varuance}, we derive expressions for the variance of the AoT and show that, in these expressions, $C({\bf{x}}-{\bf{x}}')$ appears at two distinct points in space and time. Therefore, a power-law behavior of $C({\bf{x}}-{\bf{x}}')$ could be  revealed in higher moments of the AoT. Interestingly,  power-law behavior is indeed exhibited by this correlator in higher dimensions~\cite{Poboiko_2024,Chahine_2024}, or even in 1D for interacting fermions with particle-number conservation~\cite{muller2025monitoredinteractingdiracfermions}. Alternatively, a power law can still emerge upon breaking the $U(1)$ symmetry,  or for free fermions with long-range hopping~\cite{M_ller_2022}. 

\section{Random unitary circuits interrupted by invertible measurements}
\label{se:random_quuantum_circuits}
In the previous sections we emphasized the local nature of the AoT. Here, we introduce random unitary gates—operations that drive the system into a highly entangled state obeying a volume law. These states are so strongly randomized that any underlying local structure is effectively erased. When measurements are incorporated, however, the AoT re-emerges. We will show that it exhibits nonanalytic behavior precisely at the MIPT, where the entanglement scaling transitions to an area-law phase.

Consider the canonical setup of MIPTs: a quantum circuit of $L$ qudits evolving under random two-qudit unitary gates, interspersed with $N$ layers of measurements (see Fig.~\ref{fig:lattice_permutatoin_spins}(a)). The circuit consists of independent Haar-random unitary gates acting on nearest-neighbor qudits, each with a local Hilbert space dimension $q$. After each layer of unitary gates, every qudit is measured. 

We will make use of an exact mapping of these models to statistical mechanics models~\cite{Bao_2020,Jian_2020}. In particular, we closely follow Ref.~\cite{Bao_2020}, where monitoring is implemented by coupling the system qudits ($S$) to a set of measurement ancillas ($M$). However, instead of implementing projective measurements with a controlled probability, we replace them with invertible measurements governed by a parameter $\alpha$, such that $\alpha \to 0$ corresponds to continuous measurement. Projective measurements are not invertible and therefore correspond to an infinite AoT, whereas the invertible measurements we use create a finite AoT, while maintaining exact solvability, and become continuous in an appropriate limit.

\subsection{Invertible Qudit Measurements}
\label{se:measurements}
We now present the invertible measurement procedure in two equivalent ways: First, by entangling a system's qudit with a new ancilla qudit each time and projectively measuring the ancilla — the actual strength of the measurement stems from the chosen amount of system–ancilla entanglement. Second, by a set of invertible effective Kraus operators acting only on the system qudits. While the analysis of the AoT is straightforward when using the Kraus operator description, the analysis of the random circuits will be carried out using the entangled-ancilla description, as in Ref.~\cite{Bao_2020}.

For each measurement, an ancilla qudit with the same local Hilbert space dimension $q$ is introduced in the equal superposition of basis states
\be
|s\rangle_m=\frac{1}{\sqrt{q}} \sum_{i=1}^q |i\rangle_m,
\ee
and then coupled to the system qudit via an entangling unitary gate
\be
\hat{R}_\alpha=\sum_{i=1}^q \hat{P}_i \otimes e^{- i \alpha \hat{Y}_i}.
\ee
Here $\hat{P}_i={}_s|i\rangle \langle i|_s$ and $\hat{Y}_i=-i
({}_m|s\rangle \langle s_\perp^{(i)}|_m -
{}_m|s_\perp^{(i)}\rangle \langle s|_m)$ is a usual Pauli $Y$ operator, where we defined the (normalized) state
\be
|s_\perp^{(i)} \rangle_m= \frac{|i\rangle_m - \cos \theta_0 |s\rangle_m}{\sin \theta_0},
\ee
which is orthogonal to $|s\rangle_m$ and lies in the subspace spanned by $|s\rangle_m$ and $|i\rangle_m$. Here $\cos \theta_0={}_m\langle i|s\rangle_m=\frac{1}{\sqrt{q}}$ and $\sin \theta_0=\sqrt{1-\frac{1}{q}}$. 

The entangling gate $\hat{R}_\alpha$ correlates the quantum state of
a system qudit with that of an ancilla qudit only if $\alpha\ne 0$. For example, after applying $\hat{R}_\alpha$ with $\alpha=\theta_0$ the ancilla qudit state becomes $| i \rangle_m$ if and only if the system qudit is in $| i \rangle_s$. Therefore a projective measurement of the ancilla qudit in the computational basis $| i \rangle_m$ 
corresponds to a projective measurement of the system qudit. However, when $\alpha\to0$ the resulting system and ancilla qudits are close to a product state, so a measurement of the ancilla 
reveals only little information about the system. This corresponds to the limit of continuous measurements. 

The corresponding Kraus operators could be found by looking at the density matrix $\rho_{SM}$
of a single system-ancilla
qudit pair, after the measurement procedure:
\be
\label{eq:rho_SM}
\rho_{SM} = \mathcal{N}_\phi \left[ \hat{R}_\alpha (\rho_s \otimes |s\rangle_m \langle s|_m) \hat{R}_\alpha^\dagger \right],
\ee
where $\mathcal{N}_\phi \left[ \rho \right]=\sum_i |i\rangle_m \langle i |_m \rho |i\rangle_m \langle i |_m$ is the dephasing channel acting on the ancilla that corresponds to measurement in the computational basis, and $\rho_s=| \psi_0 \rangle_s \langle \psi_0|_s$ is the density matrix of the system before the measurement. Making use of the matrix element
\bea
\label{eq:matrix_element1}
{}_m\langle j|s_\perp^{(i)}\rangle_m &=&
\begin{cases}
-\frac{\cos^2 \theta_0}{\sin \theta_0} & \text{if } j \ne i \\
\sin \theta_0 & \text{if } j=i
\end{cases} \nonumber \\
&=&-\frac{1}{\sqrt{q(q-1)}}+\delta_{ij}\sqrt{\frac{q}{q-1}},
\eea
we find the effective Krauss operators, acting only on the system subspace, to be
\bea
\label{eq:matrix_element}
{}_m \langle j | \hat{R}_\alpha | s\rangle_m&=&{}_m \langle j | \sum_{i=1}^q \hat{P}_i \left( \cos \alpha |s\rangle_m +  
\sin \alpha |s_\perp^{(i)} \rangle_m \right) \nonumber \\
&=&c_0 \hat{I} + c_1 \hat{P}_j,
\eea
where $c_0=\frac{\cos \alpha }{\sqrt{q}}
-\frac{\sin \alpha }{\sqrt{q(q-1)}}$ and $c_1=
\sin \alpha \sqrt{\frac{q}{q-1}}$. 
This can be written in a standard form as
\be 
\label{eq:Kraus_qudit}
A_j=\frac{1}{c_+} \operatorname{diag}(1, 1, \cdots, \underbrace{a}_{j\text{-th}}, \cdots, 1),
\ee
satisfying $\sum_{j=1}^q A_j^\dagger A_j=\mathbb{I}$ with $a=1+\epsilon=\frac{1+\sqrt{q-1}\tan \alpha }{1-\frac{\tan \alpha}{\sqrt{q-1}} }$ and normalization factor  $c_+^2=(q-1)+a^2$. 
These Kraus operators give a qudit generalization of the above qubit case and the measurement becomes continuous in the limit $\epsilon \to 0$ (same as $\alpha\to0$). 
Let us emphasize that the resulting state of $S$ and $M$ is no longer pure due to the presence of the dephasing channel, which in practice is obtained by entangling the ancillas with yet another bath.

\subsection{Arrow of time with  qudit measurements}
\label{se:AoT_overview}
In this section we analyze the arrow of time that is created in a general quantum circuit measured by the Kraus operators that were just introduced (Eq.~(\ref{eq:Kraus_qudit})). We find an important relation between the AoT and the Shannon entropy of the system + ancilla density matrix. This relation will be central in the following sections in which we prove a non-analytic behavior of the average AoT in the critical point of the MIPT. Moreover, it will be made clear that the local nature of the measurements leads to an AoT that depends exclusively on local observables. Still, we argue, even such a local quantity is sensitive to the entanglement transition induced by the measurements. Qualitatively, this could be seen as the effect of volume-law vs. area-law entanglement on the statistics of the local observables, but the rigorous proof will require a more careful analytical work.

Although we will eventually discuss random circuits, our analysis assumes a specific realization of such a circuit at each time. This means that over the course of any quantum trajectory the randomness only comes from the randomness of the measurements. The time-reversal of the unitary gates is deterministic and so they do not contribute directly to the AoT. 

To define the AoT, let us introduce the Kraus operators that invert the dynamics of the operators in Eq.~(\ref{eq:Kraus_qudit}), 
\be 
\tilde{A}_j=\frac{1}{c_-} \operatorname{diag}(a,\cdots,a,\underbrace{1}_{j\text{-th}},a,\cdots,a),
\ee
satisfying $\sum_{j=1}^q \tilde{A}_j^\dagger \tilde{A}_j=\mathbb{I}$ with $c_-^2=a^2(q-1)+1$.  For a given $A_j$, the operator that reverses the dynamics $\tilde{A}_{\tilde{j}}$ has $\tilde{j}=j$. 
This set of Krauss operators for $q>2$ is not built from the microscopic notion of time reversal for the qudits, but rather from an ad hoc approach of reversing the forward dynamics together with normalization. Nevertheless, every possible choice of reversal operators could differ only in the weight $c_-$, since they must effectively reverse each of the operators $A_j$. Such a difference, as we shall see in Eq.~(\ref{eq:Q_by_p_m}), can change the AoT only by a constant. Moreover, for $q=2$ the $\tilde{A}_j$'s reduce to the  microscopic time reversal construction of the previous section (up to a change of  notation).

Let us start with considering a particular trajectory through a particular circuit. A quantum trajectory corresponds to a set of measurement outcomes $m=\{ j_{x,\tau}\}$. It is useful to combine the concatenated product of unitary gates and Kraus operators into a single operator $A_m$ (satisfying $\sum_m A_m^\dagger A_m=\mathbb{I}$, see Sec. \ref{se:definition}). The probability of the trajectory is then $p_m={}_s\langle \psi_0 | A_m^\dagger A_m | \psi_0 \rangle_s$. Similarly, $\tilde{A}_m$ is the Kraus operator implementing the time-reversal of the trajectory (measurements and unitaries). 
For a given $A_m$, the operator that reverses the dynamics $\tilde{A}_{\tilde{m}}$ has $\tilde{m}=m$. 

The probability of occurrence of the backward trajectory is obtained by first evolving the state and normalizing it, $|\psi_0 \rangle \to |\psi_f\rangle= \frac{A_m | \psi_0 \rangle}{\sqrt{P_m}}$, and then computing the reversal probability $\langle \psi_f | \tilde{A}_m^\dagger \tilde{A}_m |\psi_f\rangle$. Then the arrow of time is
\be
\label{eq:Q_m}
 Q_m = \log \frac{p_m}{\tilde{p}_m} = \log \frac{p_m^2}{{}_s\langle \psi_0 |A_m^\dagger \tilde{A}_m^\dagger \tilde{A}_m A_m | \psi_0 \rangle_s}.
\ee
Using the explicit form of the Kraus operators and their time reversal, we see that 
\be
A_m^\dagger \tilde{A}_m^\dagger\tilde{A}_m A_m=
\left( \frac{a^2}{c_+^2c_-^2}\right)^{L N},
\ee
which is a constant independent of the quantum trajectory $m$. Thus the denominator in Eq.~(\ref{eq:Q_m})  is a constant factor for all quantum trajectories. This denominator normalizes the AoT to be 0 in the no-measurement limit $a=1$, $\epsilon=0$, where the forward probability for each outcome is equally $\frac{1}{q}$ and so the numerator is also constant over all trajectories.

\begin{figure}
\centering
\includegraphics[width=\columnwidth]{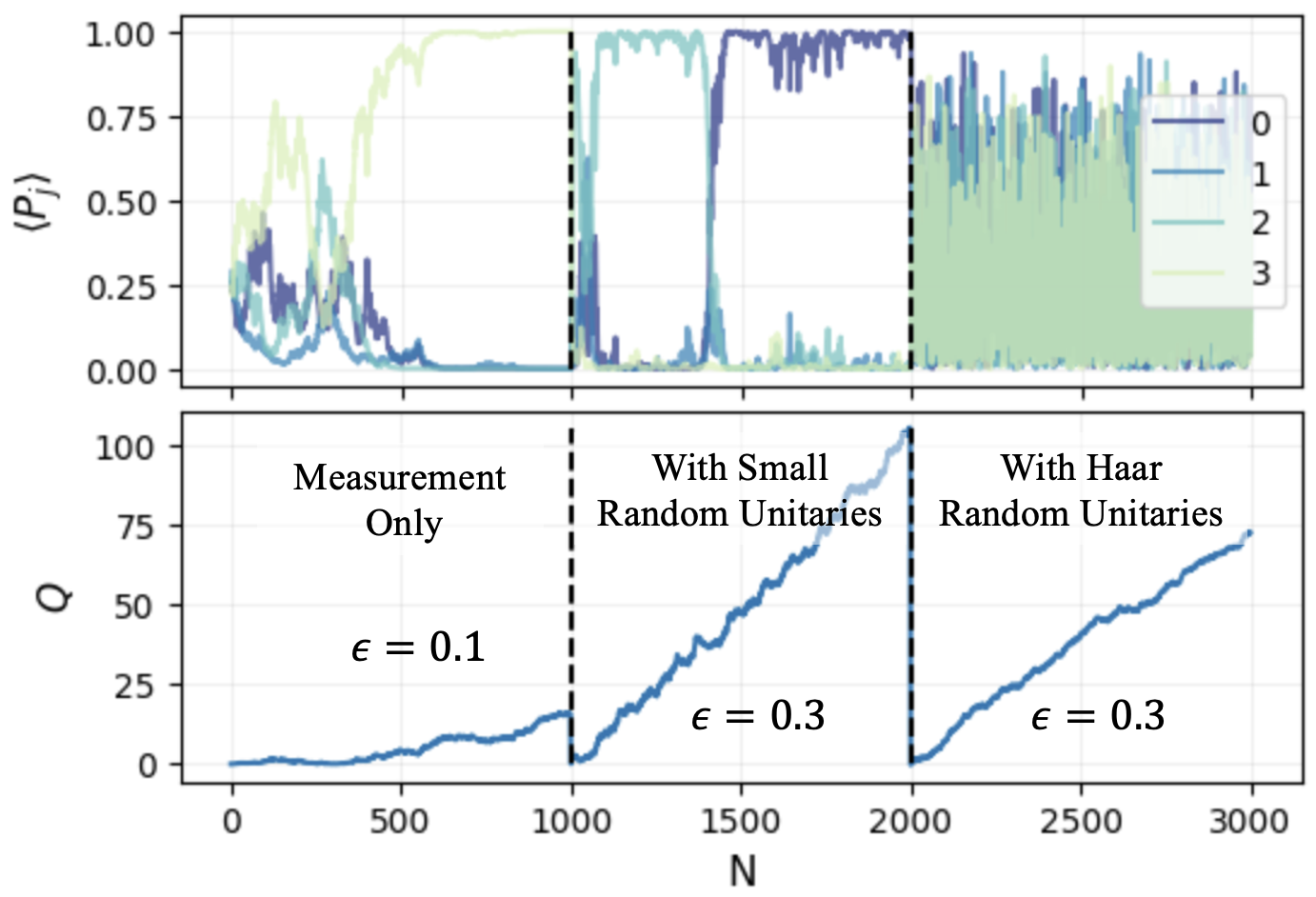}
\caption{Example trajectories of a single qudit with $q=4$ states under the Kraus operators Eq.~(\ref{eq:Kraus_qudit}) and 3 different dynamical models. Left: measurement only; 
Middle: measurements 
and random unitaries close to identity, $\exp(i\Omega H)$ for $\Omega=0.05$ and $H$ having eigenvalues $\lambda_i\le1$; Right: measurements 
and Haar-random unitaries. Each sub-trajectory is initialized in the ``equator" state $|s\rangle$. Plotted are (top) $\langle \psi_\tau|\hat{P}_j | \psi_\tau \rangle$ and (bottom) AoT. We can see a collapse to a single state in the measurement-only trajectory, 
with occasional jumps when small perturbations are introduced, and completely random expectation values when Haar-unitaries are used. This shows qualitatively that the AoT steadily increase in the usual case of any of those dynamics, with slope that is higher the stronger the measurement and the closer the state is to one of the measurement eigen-states - the AoT is smaller in the Haar-random case since it is forced to stay far from the measurement eigen-states. Importantly, as we further elaborate in Sec.~\ref{se:interpret}, we notice that $\langle \hat{P}_j \rangle$ does not distribute uniformly in the Haar-random case, see Fig.~\ref{fig:P_j_dist}.
}
\label{fig:qudit}
\end{figure}

Let us denote this constant contribution to the arrow of time by $Q_0$,
\bea
\label{eq:Q0}
Q_0 &=& - \log \left(\frac{a^2}{c_+^2 c_-^2} \right)^{LN} \nonumber \\
&=&-NL \log \frac{a^2}{[q-1+a^2][(q-1)a^2+1]} ,
\eea
leading to the simple expression for the AoT,
\be
\label{eq:Q_by_p_m}
Q_m = Q_0 + 2\log p_m,
\ee
in which the main quantity is the trajectory probability $p_m$. 
The AoT approaches the maximal value $Q_0$ only when a single trajectory is possible ($p_m=1$). While this is never the case, it implies that when the state is close to a measurement eigen-state ($\langle \hat{P}_j\rangle\approx1$ for some $j$), the few very probable trajectories that stay close to that eigen-state are the most irreversible. On the other hand when many equally possible trajectories exist, such as when starting at the ``equator" state $|s\rangle$ ($\langle \hat{P}_j\rangle\approx \frac{1}{q}$ for all $j$), the arrow of time increases much slower. Fig.~\ref{fig:qudit} (left panels) illustrates both of these scenarios - the state is initialized at the equator, and so the AoT increases slowly; Then the state collapses to $\langle \hat{P}_3\rangle\approx1$ and the AoT increases much further. In Fig.~\ref{fig:qudit} (middle panels) random perturbations try to knock the state out of the measurement eigen-states, and the now stronger measurements keep the state pinned, producing much higher irreversibility.

This dependence of the AoT on the expectation values $\langle \hat{P}_j\rangle$ can be clearly seen by writing $p_m$ explicitly. Using Eq.~(\ref{eq:Kraus_qudit}) to take the measurement operators as $A_j = \frac{1}{c_+}(\mathbb{I}+(a-1)\hat{P}_j)$ we obtain

\be
\label{eq:p_m_locality}
p_m = \left(\frac{1}{c_+^2}\right)^{LN}\prod_{x=1}^L\prod_{\tau=1}^N (1+(a^2-1)\langle \hat{P}_{j_{x,\tau}} \rangle).
\ee
Since $a^2-1>0$, it is immediately clear that the higher the expectation values of the records $\{j_{x,\tau}\}$, the higher the probability of the trajectory and the higher is the increase in the AoT. This is the direct result of the locality of the measurements which allows to write $p_m$ as a product of $L$ seemingly-independent  probabilities in each time step (the implicit dependence due to entanglement shows up through the post-measurement values of $\langle \hat{P}_{j_{x,\tau}} \rangle$).  

Noticing this intimate dependence of the AoT on local expectation values, it is not at all obvious how could it show any signature of an entanglement transition, in particular in random circuits, where those expectation values get completely randomized between measurement cycles. See Fig.~\ref{fig:qudit} (right) for example. An interesting subtlety of Eq.~(\ref{eq:p_m_locality}) in that regard is the implied dependence between expectation values in the same time $\tau$ at different positions. Assuming the order of measurements is from the qudit at $x=1$, measuring one by one up until the qudit at $x=L$, then any qudit in position $i$, could be affected by the measurement of the qudits at $x<i$ through entanglement. But after all, the qudit $i$ had completely random local expectation values before the measurements of its preceding qudits, and it has completely random expectation values afterwards, so this subtlety seems to be trivially resolved.

The actual clue lies in the probability distribution of those expectation values. As seen in Fig.~\ref{fig:qudit} (right), $\langle \hat{P}_j\rangle$ does not distribute uniformly. In general, the shape of this distribution depends on the dimension of the Hilbert space over which the state is randomized.
As it turns out, the higher the dimension, the narrower the distribution of $\langle \hat{P}_j\rangle$ around $\frac{1}{q}$. In that sense, the entanglement transition from volume-law to area-law could be seen as nothing but the sudden reduction of the available Hilbert space from a $\sim q^L$ dimensional space to a product of $\sim q$ dimensional subspaces. This then shows up as higher possible $\langle \hat{P}_j\rangle$ values, and thus a higher arrow of time. While this is only a qualitative argument, it will be made more concrete with a few examples in Sec.~\ref{se:interpret}. And nevertheless, its main purpose is to help make sense of the analytical results of the following sections.

Proceeding to the trajectory-averaged AoT, we can now see an important relation,
\be
\label{eq:Stinesprint_dilation}
\overline{Q}= \sum_m p_m Q_m = Q_0 + 2 \sum_m p_m \log p_m = Q_0 - 2 S_{SM}.
\ee
Here $S_{SM}=-\sum_m p_m \log p_m$ is 
the Shannon entropy  of the measurement outcomes. We emphasize here that the density matrix of the system and ancilla qudits, $\rho_{SM}$, is mixed due to the dephasing channel measuring the ancillas (see Eq.~(\ref{eq:rho_SM})). One can purify this channel by including an additional bath, such that the entanglement of the system and ancilla qudits with the bath equals the Shannon entropy of the measurement distribution function, 
\be
\label{eq:sss}
-\sum_m p_m \log p_m=-\ {\rm{Tr}} \rho_{SM} \log \rho_{SM}=S_{SM}.
\ee

Finally, $S_{SM}$ can be obtained from the replica limit
\be
\label{eq:replica}S_{SM}=\lim_{n \to 1} \frac{1}{1-n}\log {\rm{Tr}} \rho_{SM}^n = -\lim_{n \to 1} \partial_n \log {\rm{Tr}} \rho_{SM}^n. 
\ee
Precisely this object, upon averaging over the random unitary gates from the Haar ensemble and for the case of projective measurements which occur with a finite probability, was  mapped to a classical statistical mechanical problem by Bao et al.~\cite{Bao_2020} and  Jian et al.~\cite{Jian_2020}. This provided an exact and analytic mapping of the MIPT to known classical statistical mechanics phase transitions. Here we will obtain a similar mapping for the invertible measurements introduced above, leading to 
the trajectory averaged and circuit averaged AoT,
\be
\langle\langle \overline{Q}\rangle \rangle = Q_0 + 2 \lim_{n \to 1} \partial_n \log Z^{(n)},
\ee
where $Z^{(n)} = \langle\langle {\rm{Tr}} \rho_{SM}^n\rangle\rangle$ is the partition function of the equivalent classical model. This mapping allows us to identify exactly the critical behavior of the AoT in appropriate limits.

Let us recall our definitions of averages: averages with respect to quantum trajectories are denoted by an over-bar, for example $\overline{ Q}=\sum_m p_m Q_m$. This should be distinguished from the usual quantum average, and from random unitary circuit (RUC) averages:
\bea
\langle O \rangle:&& {\rm{quantum~average}}, \nonumber \\
\overline{ O}:&& {\rm{quantum~trajectory~average}}, \nonumber \\
\langle \langle O \rangle \rangle: &&{\rm{RUC~average}}.
\eea

We refer the impatient reader to Sec.~\ref{se:interpret}, where the results of the following extended sequence of exact mappings—which, in essence and aside from our required modifications, already appear in Refs.~\cite{Bao_2020,Jian_2020}—are discussed from a physical perspective, leading to a clear picture of the distinct role of the AoT in MIPTs.

\subsection{AoT as a  classical partition function}
\begin{figure*}[t]
\centering
\includegraphics[width=\textwidth]{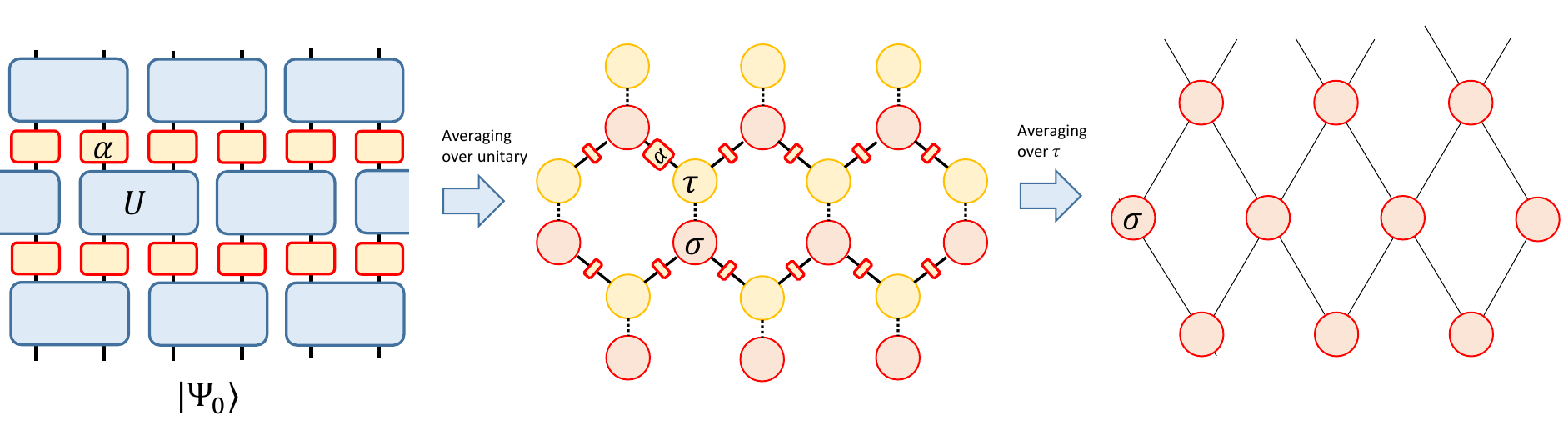}
\caption{Reproduced  from Ref.~\cite{Bao_2020} for clarity, with small modifications. (a) Random unitary circuit of $L$ qudits with local Hilbert space dimension $q$, interrupted by  measurements which we modify to be invertible, as parameterized by $\alpha$ - see Eq.~(\ref{eq:rho_SM}).
(b) The density matrix of $n$ copies of both the system qudits and measuring ancilla qudits, ${\rm{Tr}} \rho_{SM}^n$, averaged over random unitary gates, maps to a statistical mechanics model on the honeycomb lattice. Each local degree of freedom $\sigma,\tau$ takes $n!$ states and corresponds to a permutation in replica space. Diagonal bonds describe measurements, which we take to be non-projective, i.e., invertible. 
(c) Averaging over the $\tau$ sublattice the model maps to a square lattice with variables $\sigma$. In the large $q$ limit we obtain an exactly solvable weight function, Eq.~(\ref{eq:Weight}), which allows us to take the $n\to 1$ replica limit and identify the MIPT with bond percolation as in the projective measurement case~\cite{Bao_2020}. 
} 
\label{fig:lattice_permutatoin_spins}
\end{figure*}

The $n$-th
moment of a density matrix evolved by a RUC can be mapped to the partition function of a classical
spin model with $n!$ states~\cite{Nahum_2017,Zhou_2019,hunterjones2019unitarydesignsstatisticalmechanics}. 
Since the $n$-th moment of the density matrix is an n-th order monomial of random unitary gate $U$ and its conjugate $U^\dagger$, by averaging over the unitary gates, one can write $ \langle \langle {\rm{Tr}} \rho_{SM}^n \rangle \rangle$   as
a sum of terms labeled by two emergent spin variables $\sigma$
and $\tau$, taking
$n!$ different values and actually being elements of the permutation group, each living on one of the two honeycomb sublattices, see  Fig.~\ref{fig:lattice_permutatoin_spins}(b).
The resulting partition function is a classical model of these permutation variables,
\be
\label{eq:rho_n_to_partition_functino}
\langle \langle {\rm{Tr}} \rho_{SM}^n \rangle \rangle=Z^{(n)}=\sum_{\{\sigma_{\bf{r}},\tau_{\bf{r}'}\}} \prod_{\langle {\bf{r}},{\bf{r}'} \rangle} W^{(n)}(\{\sigma_{\bf{r}} ,\tau_{\bf{r}'}\}),
\ee
where $W^{(n)}(\{\sigma_{\bf{r}} ,\tau_{\bf{r}'}\})$ are the weights associated with a given spin configuration $\{\sigma_{\bf{r}} ,\tau_{\bf{r}'}\}$. Those weights are functions of nearest neighbor spin variables
at $\langle {\bf{r}},{\bf{r}'} \rangle$ in the lattice given by $W^{(n)}=w^{(n)}_d$ or $w^{(n)}_g$ depending on the orientation of the bond. The vertical bonds describe the ensemble averaged Haar random two-qubit gates, and are known
Weingarten functions~\cite{collins2003moments}. 
On the other hand, the diagonal bonds $w^{(n)}_d(\sigma,\tau)$ result from measurements, as introduced in Refs.~\cite{Bao_2020,Jian_2020} for the case of projective measurements. Here we will derive $w^{(n)}_d(\sigma,\tau)$ for our invertible measurements. Relegating the derivation to Appendix~\ref{appendix:w_d}, let us directly state the result: Like $w^{(n)}_g(\sigma,\tau)$, also $w^{(n)}_d(\sigma,\tau)$ depends only on the relative permutation  
 of the neighboring sites $\xi=\tau^{-1} \sigma$. It can be expressed solely in terms of the cycle decomposition of $\xi$: one can decompose each permutation into $k$ cycles of lengths $\{n_1,n_2,\dots,n_k\}$, such that $\sum_{i=1}^k n_i =n$. Here $k=\# {\rm{cycles}}(\tau^{-1} \sigma)$. For example the identity consists of $n$ cycles of length 1 $\{n_i=1|i=1,\dots,n \}$ while the cyclic permutation consists of a single cycle of length $n_1=n$. We find that 
\bea
\label{eq:w_d}
w^{(n)}_d(\sigma,\tau)&=&q  \prod_{i=1}^{\# {\rm{cycles}}(\tau^{-1} \sigma)} [\left(\frac{\cos \alpha+\sin \alpha \sqrt{q-1}}{\sqrt{q}} \right)^{2n_i} 
 \\
&&+(q-1)\left(\frac{\cos \alpha -(\sin \alpha) / \sqrt{q-1}}{\sqrt{q}} \right)^{2n_i} ]. \nonumber
\eea
It is easy to check that in the projective measurement limit, $\alpha=\theta_0$, this gives $w^{(n)}_d(\sigma,\tau)=q $, in exact agreement with Ref.~\cite{Bao_2020}. Since in this limit $w^{(n)}_d(\sigma,\tau)$ is independent of $\sigma$ and $\tau$, the statistical mechanics model is in the complete paramegnetic phase with $n!$ equally probable states per site~\cite{Bao_2020,Jian_2020}.

In the no-measurement limit $\alpha=0$ we obtain
\bea
\label{wdalpha_0}
w^{(n)}_d(\sigma,\tau)_{\alpha=0}=q^{\# {\rm{cycles}}(\tau^{-1} \sigma)+1-n}.
\eea
In this limit any misalignment of neighboring permutations is penalized by a factor $1/q$. This expression coincides up to an unimportant overall factor with Ref.~\cite{Bao_2020}. (The no-measurement case $\alpha=0$ in our model is different from Ref.~\cite{Bao_2020}). Irrespectively, this interaction forces a complete ferromagnetic alignment of the permutations.

Having identified a ferromagnetic phase in the weak-measurement limit and a paramagnetic phase in the strong-measurement limit, it is natural to infer a phase transition at a finite value of $\alpha$, which characterizes the MIPT in our model.

We expect this transition to occur for arbitrary $q$. This can be proven in particular cases. In Ref.~\cite{Bao_2020}, the corresponding statistical mechanics model was mapped onto an Ising model in the special case $n = 2$, for any $q \ge 2$. A similar analysis can be performed for our invertible measurements allowing to address the transition in the qubit case $q=2$. However we will not follow this here. Instead, since our primary interest lies in the limit $n \to 1$, we proceed to the large $q$ limit.

\subsection{Exact solution at large $q$}
Several simplifications occur for large $q$. As used in Ref.~\cite{Bao_2020,Jian_2020}, the large $q$ limit of the Weingarten function is
\be
\label{eq:w_g}
w^{(n)}_g(\sigma,\tau)_{q \to \infty}=\begin{cases}
1/q^{2n} & \text{if } \sigma =\tau \\
\mathcal{O}(1/q^{2n+1}) & \text{if } \sigma \ne \tau
\end{cases}.
\ee
It thus follows that $\sigma=\tau$ across vertical bonds of the honeycomb lattice, resulting in a square lattice, see Fig.~\ref{fig:lattice_permutatoin_spins}(c). The interaction between spins on neighboring sites $\langle x,y \rangle$ on the square lattice are determined by $w^{(n)}_d(\sigma_x,\sigma_{y})$.

In the limit of large $q$ we have $(k=\# {\rm{cycles}}(\sigma_{y}^{-1} \sigma_x))$
\bea
w^{(n)}_d(\sigma_{x},\sigma_{y})_{q \to \infty}&=&q  \prod_{i=1}^{k} \left[\sin^{2n_i} \alpha + \frac{\cos^{2n_i} \alpha}{q^{n_i-1}}  \right]. 
\eea
Noticing that every $1-$cycle ($n_i=1$) contributes a factor of unity to this product,  in the limit of large $q$ this simplifies to
\bea
w^{(n)}_d(\sigma_x,\sigma_{y})_{q \to \infty}&=&q  (\sin^2 \alpha)^{n - n_{fp}(\sigma_{y}^{-1} \sigma_x)}, 
\eea
where $n_{fp}(\sigma^{-1}_{y} \sigma_x)$ is the number of 1-cycles i.e. the number of fixed points of the permutation $\sigma_{y}^{-1} \sigma_x$.

In addition, in the limit of large $q$ we have $a^2-1 = q \tan^2 \alpha + \mathcal{O}(1)$ from which \be
Q_0 =2  N L (\log q - \log \cos \alpha).
\ee
It is convenient to combine $Q_0$ with the factors of $1/q^{2n}$ appearing in $w_{g}^{(n)}$ and the factors of $q$ in $w_d^{(n)}$. Since on the honeycomb lattice
\bea
\#~{\rm{vertical~bonds}}=\#~\sigma'{\rm{s}} = \frac{NL}{2}, \nonumber \\
\#~{\rm{diagonal~bonds}}=NL,
\eea
we can write
\bea
\label{eq:partial_n}
\langle \langle \overline{Q} \rangle\rangle  &=&Q_0+ 2\lim_{n \to 1} \partial_n \log Z^{(n)} \nonumber \\
&=&   N L \log \frac{q^2}{\cos^2 \alpha}  + 2\lim_{n \to 1} \partial_n \log [\left( \frac{1}{q^{2n}}\right)^{\frac{NL}{2}}  q^{NL}  \tilde{Z}^{(n)}]\nonumber \\
&=& 2\lim_{n \to 1} \partial_n \log \tilde{Z}^{(n)}-  N L \log \cos^2 \alpha.
\eea
Here the partition function $\tilde{Z}^{(n)}$ is defined on the square lattice,
\be
\label{eq:Ztilde}
\tilde{Z}^{(n)}=\sum_{\{ \sigma_x \}} \prod_{\langle x,y \rangle} W(\sigma_x,\sigma_{y}),
\ee
with weights 
\be
\label{eq:Weight}
W(\sigma_x ,\sigma_{y})=(\sin^2 \alpha)^{n-n_{fp}(\sigma_x \sigma_{y}^{-1})},
\ee
where $n_{fp}(\sigma_x \sigma_{y}^{-1})=\# {\rm{1-cycles}}(\sigma_{y}^{-1} \sigma_x)$.

Let us make consistency checks. In the no-measurement limit, $\alpha \to 0$, the weights force $n_{fp}=n$, namely a complete ferromagnetic alignment of the permutations. The boundary condition~\cite{Bao_2020,Jian_2020} completely fixes all permutations to a single value so that $\tilde{Z}^{(n)}=1$ and the AoT vanishes, as required.

In the projective measurement limit $\alpha \to \pi/2$, the statistical mechanics model is in the complete paramegnetic phase with $n!$ equally probable states per site. Thus $\tilde{Z}^{n}=(n!)^{NL}$. In total $Q=2NL(1-\gamma)-2  N L \log \cos \alpha$ where $\gamma$ is the Euler constant and we used $n!=\Gamma(n+1)$ to take the replica limit. As required, the AoT diverges in the projective measurement limit due to the term $-2  N L \log \cos \alpha$.

\subsection{Mapping to bond-percolation}
\label{subsec:percolation}
In Appendix~\ref{appendix:FK} we demonstrate an exact mapping of our model Eq.~(\ref{eq:Ztilde}) and (\ref{eq:Weight})
to percolation on a square lattice with bond occupation probability $p=\cos^2 \alpha$. We apply a modified Fortuin-Kasteleyn  cluster mapping~\cite{fortuin1972random} to our model, generalizing the case of  Ref.~\cite{Bao_2020} where the weights functions at each link $\langle x, y\rangle$ take only two values depending on whether $\sigma_x=\sigma_y$ or $\sigma_x\neq\sigma_y$.

\begin{figure}
\centering
\includegraphics[width=\columnwidth]{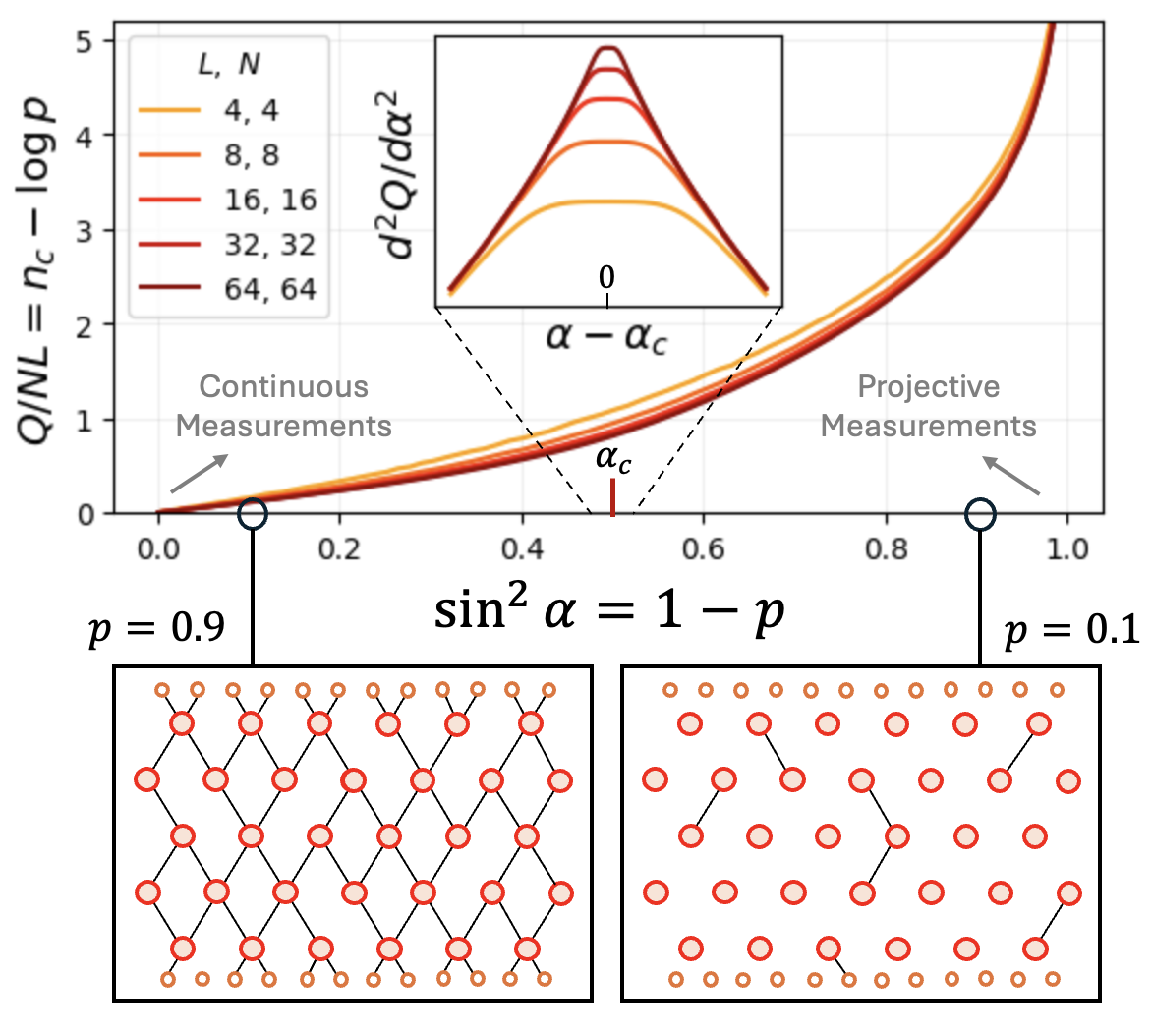}
\caption{AoT in a random  circuit of $L$ qudits with a large  Hilbert space dimension $q \gg 1$, interrupted by $N$ measurements 
parameterized by $\alpha$, where $\alpha \to 0$ corresponds to continuous measurements  and $\alpha \to \pi/2$ to  projective measurements. Results are obtained by counting clusters on small systems using Eq.~(\ref{eq:Q_RUC}), based on our exact mapping to bond percolation.
The inset schematically displays the scaling $|p-p_c|^{2/3}$ emerging at large system sizes (we schematically display a  finite size scaling plot based on a universal scaling function). We refer the reader to Ref.~~\cite{Mertens_2017} for 
large scale percolation results.
}
\label{fig:n_p}
\end{figure}

We now discuss the AoT within the mapping to percolation. One of the  fundamental quantities  in percolation is simply the number of clusters
per site $n_c(p) =\frac{N_{cluster}}{N_{site}}$. It is directly related to the more familiar number of clusters of size $s$ per volume, $n_s$~\cite{stauffer2018introduction}, via $n_c(p)=\sum_s n_s$. To acquire basic intuition on this quantity, at $p=0$ each site is a cluster so that $n_c(p=0)=1$. At finite and small $p$ there is a finite probability for neighboring sites to join into clusters, yielding $n_c(p) \approx 1-2p$, $(p \ll 1)$. Eventually  near $p \to 1$ there is one big cluster so that $n_c(p) =0$ in the thermodynamic limit. 

Using our generalized  Fortuin-Kasteleyn  cluster mapping in Appendix~\ref{appendix:FK}, we find that Eq.~(\ref{eq:partial_n}) becomes
\be
\label{eq:Q_RUC}
\langle \langle \overline{Q} \rangle\rangle  = 2 
N_{cluster} -  N L \log p,
\ee
or equivalently using $N_{site}\sim \frac{NL}{2}$, we have $\frac{Q}{NL}=
n_c-\log p$. In deriving this result the limit $n \to 1$ was taken. 

The AoT based on this exact mapping to bond percolation is plotted in Fig.~\ref{fig:n_p} versus measurement strength all the way from weak to projective measurement. It is seen to converge to the thermodynamic limit quite quickly. 

The percolation transition, which is known to be at $p_c=1/2$ for bond percolation on the square lattice, giving $\alpha_c=\pi/4$, shows up as a nonanalytic behavior in this function. Near $p_c$ one can expand~\cite{domb1976mean}
\be
n_c(p)=A_0+B_0(p - p_c)+C_0(p-p_c)^2+ \mathcal{A}^\pm |p-p_c|^{2-\alpha},
\ee
where the first three terms represent the analytical part
of $n_c(p)$, and the last term represents the singular part. In two dimensions, the critical exponent $\alpha=\alpha'=2-d \nu$ has the universal value $-2/3$ (not to be confused with the measurement strength parameter $\alpha$) and $\mathcal{A}^+=\mathcal{A}^-$. This yields a nonanalytic behavior in the AoT at the MIPT
\be
\label{eq:p8_3}
{\rm{AoT}} \sim |\alpha-\alpha_c|^{8/3},
\ee
whose third derivative diverges at the transition in the thermodynamic limit as illustrated in Fig.~\ref{fig:n_p}. The nonanalytic contribution has the finite-size scaling form
$n_c(p)=L^{-d} f(z)$, where $z=(p-p_c)L^{1/\nu}$.
The scaling function $f(z)$ is analytic near the origin, while for $|z|\gg 1$
it behaves as $f(z)\sim \mathcal{A}^{\pm} |z|^{2-\alpha}$.
To illustrate this crossover schematically, we plot in the inset of
Fig.~\ref{fig:n_p} a simple interpolating form, defined via
$f''(z)=(1+|z|^{b})^{\alpha/b}$ (with $b=4$), which reproduces the correct
limiting behaviors but is not intended to represent the exact scaling function.
For a recent study of percolation demonstrating this singular behavior using
large-scale numerical simulations, see Ref.~\cite{Mertens_2017}.

\subsection{Influence of entanglement transitions on the arrow of time}
\label{se:interpret}
In the last subsections, that were notoriously technical, we established the mapping from $n$ replicas of the average RUC model to a classical lattice model. The trajectory-averaged and circuit-averaged AoT, $\langle\langle \overline{Q}\rangle \rangle$, was found to be related to the free energy of the lattice at the replica number limit $n\to1$ (see Eq.~(\ref{eq:partial_n})). This classical model was shown to have a ferromagnetic-paramagnetic transition at the equivalent of a finite measurement strength $\alpha$, which points at a non-analytic transition in the AoT in the appropriate replica limit. 
For large qudit dimension $q$, this replica limit $n\to1$ was solved exactly, leading to a 2D percolation model through which the critical measurement strength $\alpha_c = \frac{\pi}{4}$ was identified together with all the critical exponents (see Eqs.~(\ref{eq:Q_RUC}-\ref{eq:p8_3})).

In this subsection, we return to the physical picture of the AoT and appeal to a simple intuition that emerges from our findings.
As was briefly discussed in Sec.~\ref{se:AoT_overview}, the AoT is a function of the quantum trajectory which only depends on local observables.
Following Eqs.~(\ref{eq:Q_by_p_m}-\ref{eq:p_m_locality}) we can write it explicitly for each trajectory $m=\{j_{x,\tau}\}$ as
\be
\label{eq:Q_locality}
Q_m = Q_0 + 2 \sum_{x=1}^L\sum_{\tau=1}^N \log{\left(\frac{1+(a^2-1)\langle \hat{P}_{j_{x,\tau}} \rangle}{q + (a^2 - 1)}\right)}.
\ee
It is then questionable whether such a local quantity could reveal any signature of an entanglement transition, especially in the presence of Haar-random unitaries that completely randomize the state on a local level.
Assuming the local expectation values at each time step
$\langle \hat{P}_j\rangle$ are truly random and independent (we shall return to this assumption shortly), the trajectory average and RUC average of the AoT could be written schematically as a single integral
\be
\label{eq:approx_AoT}
\langle\langle \bar{Q}\rangle \rangle \sim \int  d \langle \hat{P}_j\rangle p(\langle \hat{P}_j\rangle) Q(\langle \hat{P}_j\rangle),
\ee
where $\langle \hat{P}_j\rangle$ refers to the set of all $j=1,\dots,q$ expectation values, which distribute identically through space and time. As just noted, it is unclear whether $Q(\langle \hat{P}_j\rangle)$ could be affected by the entanglement structure, and how it can change across the MIPT. In contrast, the probability distribution $p(\langle \hat{P}_j\rangle)$ is strongly affected by the MIPT, and is the key for understanding the non-analytic transition in the AoT, as we shall now explain in detail.

Given that there is an entanglement transition, as already established, we will approximate the quantum state in the volume-law phase as a Haar-random state in the $\mathcal{O}(L)$ qudit Hilbert space. Similarly, the quantum state in the area-law phase will be approximated as a product state of local Haar-random states over $\mathcal{O}(1)$ qudits. This crude approximation is based on the assumption that when the state is volume-law entangled, it performs a ``random walk" over the entire $L$ qudit Hilbert space, and covers it uniformly over long times, even though each time step consists of only 2-qudit unitary gates. In the strong-measurement phase, the measurements pin the state to the local Hilbert space of each qudit, with the 2-qudit unitaries being able to entangle only neighboring or nearly-neighboring qudits. Even though there could be some dependence between subsequent time steps, over long times this dependence will have a vanishingly small effect on the average AoT, as long as the coverage of the available Hilbert space by the state will be uniform, thus justifying Eq.~(\ref{eq:approx_AoT}).

\begin{figure}
\centering
\includegraphics[width=0.9\columnwidth]{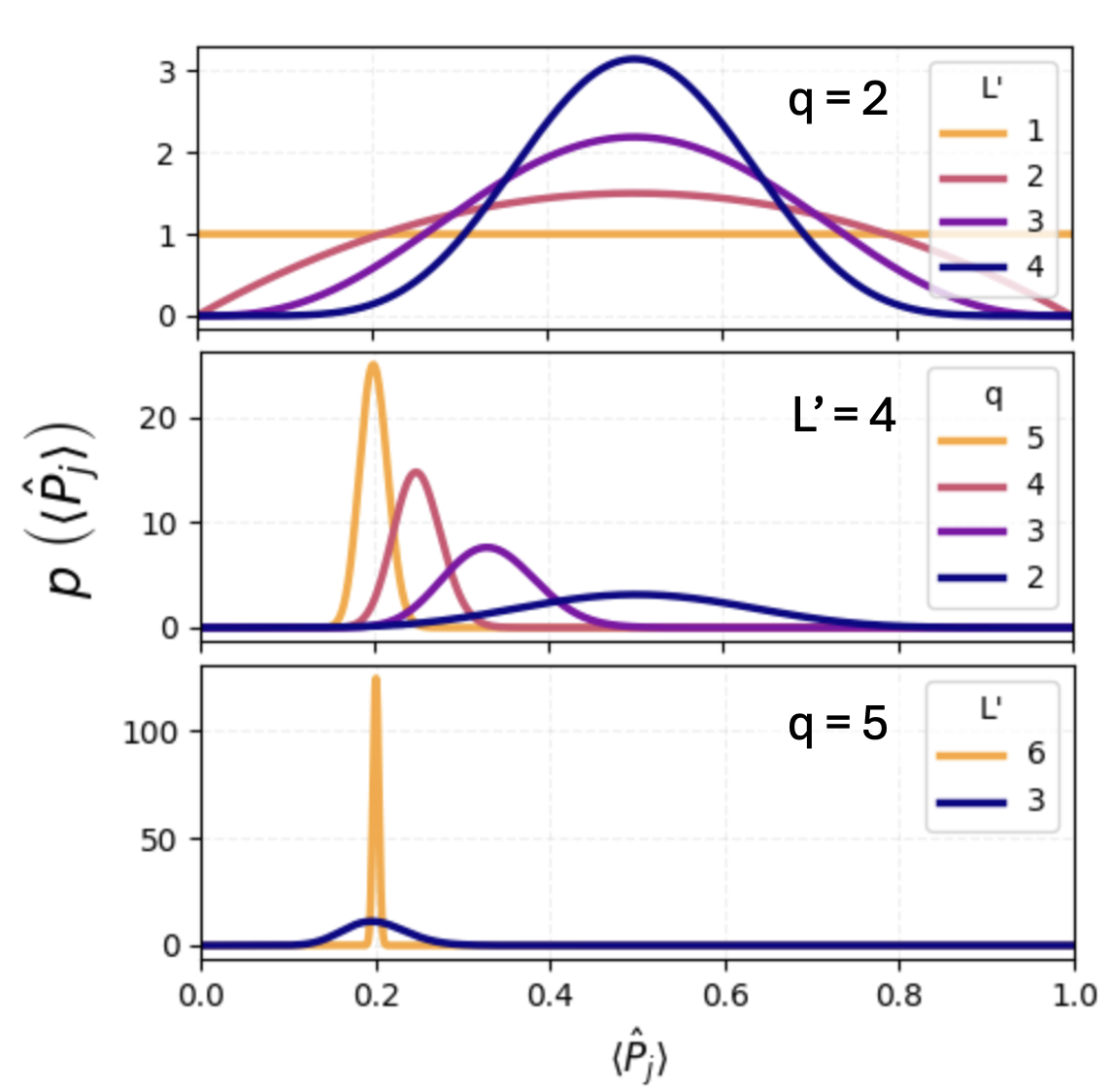}
\caption{Distributions of the local expectation values $\langle \hat{P}_j \rangle$ over a family of Haar-random states of $L'$ qudits of dimension $q$. These distributions illustrate the following points: (top) qubits ($q=2$), that are known to have uniform distribution when $L'=1$, have narrower distribution for random states of larger number of qubits; (center) for the same number of qudits, increasing the local dimension $q$ shifts the peak to be centered around $1/q$; (bottom) for large $q$ the distribution changes dramatically between small and large $L'$, illustrating the difference between the volume-law phase where $L'=\mathcal{O}(L)$ and the area-law phase where $L'=\mathcal{O}(1)$. Plotted are the Beta distributions ${\rm{Beta}}(q^{L'-1}, q^{L'}-q^{L'-1})$ for which the variance goes like $1/q^{L'+1}$ for large $q$.}
\label{fig:P_j_dist}
\end{figure}

Let us denote by $L'$ this effective number of entangled qudits, which goes from $\mathcal{O}(L)$ to $\mathcal{O}(1)$ at the MIPT. The difference in the AoT in the two phases is then encapsulated in the corresponding distinct behaviors of $p(\langle \hat{P}_j\rangle)$ for different $L'$. 
For a product state of single qubit states $L'=1$, a Haar-random state means $\langle \hat{P}_j\rangle$ distribute uniformly as shown in  Fig.~\ref{fig:P_j_dist} (top). This is simply a state of random pure Bloch-vectors, so the probability distribution of any Pauli operators $\langle\sigma_i\rangle$ is uniform over $[-1,1]$. 
However, when randomizing a product state of multiple qubits $L'>1$, the resulting state is usually entangled, and the reduced density matrix of each degree of freedom becomes mixed, leading to a narrower distribution around the ``equator" state. This narrowing increase as $L'$ increase, see Fig.~\ref{fig:P_j_dist} (top), eventually leading to a delta function around the maximally mixed local state at $L'\to\infty$. 

In qubits it is a well-known phenomena seen through the shrinking of the local Bloch-vectors in entangled states. For qudits, the dimension $q$ also alters the distribution. A maximally mixed state of $q$ dimensions has a weight of $1/q$ for each possible state, then similarly the $\langle \hat{P}_j\rangle$ distribution shifts such that its peak is over $1/q$, see Fig.~\ref{fig:P_j_dist} (center).

Over the MIPT, we consider a large yet finite $q$, then the variance of those distribution functions  is ${\rm{Var}}(\langle \hat{P}_j\rangle) \sim \frac{1}{q^{1+L'}}$. Meaning that the distribution changes from having a finite width when $L'=\mathcal{O}(1)$ in the strong-measurement phase, to being a sharp delta function when $L'=\mathcal{O}(L)\to\infty$ in the volume-law entanglement phase, see Fig.~\ref{fig:P_j_dist} (bottom).

Going back to Eq.~(\ref{eq:approx_AoT}), we now see how a sharp change in the entanglement leads to a sharp change in the variance of $p(\langle \hat{P}_j\rangle)$ which finally, since $Q(\langle \hat{P}_j\rangle)$ is far from being a linear function, induce a significant change in the average AoT $\langle \langle \bar{Q} \rangle \rangle$, as shown analytically in the previous sections.

\section{Summary and outlook}
\label{se:summary}
In this work, we applied a microscopic, time-reversal--based definition of the arrow of time to monitored quantum systems undergoing measurement-induced phase transitions. We illustrated this perspective through a series of settings. The simplest case, involving no-click monitoring dynamics, demonstrates the idea at the level of a single quantum trajectory. We then extended the analysis to the generic situation with many quantum trajectories, showing that the measurement-induced arrow of time is directly related to a local correlation function 
which is nonlinear in the density matrix.

To demonstrate unambiguously that the arrow of time exhibits nonanalytic behavior at the measurement-induced phase transition, we revisited random quantum circuits composed of qubits with a large local Hilbert space. We replaced projective measurements with invertible ones, allowing for a more controlled analysis. As in the well-studied case of projective measurements, we found that in the limit \( q \to \infty \), the universality class of the transition matches that of bond percolation on the two-dimensional square lattice. This correspondence enabled us to determine the critical properties of the arrow of time, which are governed by the thermodynamic critical exponent associated with the specific heat, \( \alpha = -2/3 \).

The arrow of time is distinguishable in several important ways from previously studied quantities that probe MIPTs. Most quantities considered so far are information-theoretic in nature, such as entanglement measures, quantum error correction and scrambling dynamics. Other probes include the channel capacity and related purification transition~\cite{Gullans_2020}, and the sensitivity of measurements to the initial state, as quantified by the Fisher information~\cite{Bao_2020}. Within the statistical mechanics mapping, all of these observables correspond to various surface phenomena associated with different \emph{boundary conditions}~\cite{Bao_2020,Jian_2020}. For example, the phase transition in the scaling of entanglement corresponds to the free energy density of a domain wall induced by a change in boundary conditions and a similar boundary condition interpretation applies to the purification transition.

In contrast, the AoT probes the \emph{bulk properties} of the theory directly. In this sense it is a true thermodynamic probe. It is also distinguished by the type of quantity it accesses in the bulk. A local order parameter for MIPTs has been identified based on the entropy of an external qubit initially entangled with a specific site~\cite{Gullans_2020micro}. By comparison, the AoT is related to the free energy of the theory, and in particular captures the specific heat exponent, $\alpha$. Its critical behavior can thus be probed from both sides of the transition, including in the disordered phase where the conventional order parameter vanishes. Thus, the AoT, serving as the free energy operator, is a useful tool to identify the underlying conformal field theory on which a lot of effort is being devoted~\cite{Zabalo_2022}.

A key result of this work is that the AoT can be directly extracted from the entropy of measurement outcomes, see Eq.~(\ref{eq:Stinesprint_dilation}). This places the AoT at an advantageous position with respect to the postselection overhead problem that arises in entanglement-based probes of measurement-induced phase transitions, where one needs to repeatedly prepare identical quantum states in order to reliably estimate expectation values. In contrast, according to  Eq.~(\ref{eq:Stinesprint_dilation}) the AoT requires no evaluation of quantum expectation values, relying instead solely on the entropy of the measurement outcomes. Yet, as we have shown, the critical behavior of the AoT shows up  only in its derivatives  with respect to the measurement rate. Thus, in practice one has to device ways to directly evaluate derivatives — which is left for future work.

Looking ahead, it is natural to explore the AoT diagnostic in non-projective measurement–based phenomena in various contexts other than quantum circuits. For example, mesoscopic realizations of systems exhibiting the Kondo effect, perturbed by the backaction of monitoring of the charge or spin of a localized electron in a quantum dot~\cite{buks1998dephasing,Sankar_2024,sankar2025backactioneffectschargedetection}, have been a fruitful test bed for thermodynamic entropy measurements~\cite{Hartman_2018,Child_2022} and stochastic thermodynamics~\cite{hofmann2017heat,barker2022experimental,han2024quantum,ma2025quantum}. Turning quantum measurement to be the main process driving the dynamics and often creating Zeno-like transitions~\cite{nakagawa2018non,lourencco2018kondo,hasegawa2111kondo,kattel2402dissipation,vanhoecke2405kondo,ma2023identifying,stefanini2025dissipative,werner2026nonequilibrium,roadmap2026roadmap} is an interesting direction to study the AoT and entropy production.

More generally, the AoT can probe irreversibility across a broad class of non-projective measurement–based phenomena, including the stabilization of topological states by non-commuting measurements~\cite{Kells_2023}, topological transitions~\cite{Gebhart_2020}, nonunitary measurement based quantum computing and imaginary-time simulators~\cite{TERASHIMA_2005,Azses_2024}, mappings of monitored circuits to statistical mechanics models~\cite{Zhu_2023,P_tz_2025}, explicit encoding–decoding quantum error correction~\cite{Turkeshi_2024} and magic phase transitions~\cite{niroula2024phase}.


\begin{acknowledgments}
ES and NH gratefully acknowledge support from the European Research Council (ERC) under the European Union Horizon 2020 research and innovation programme under grant agreement No. 951541. We thank fruitful discussions with Sreenath Manikandan, Dganit Meidan, Serge Rosenblum, Jonathan Ruhman and Jagannath Sutradhar.
\end{acknowledgments}

\begin{appendix}

\section{AoT in non-hermitian systems}
\label{appendix:non-hermitian}

We discuss a general relation for the arrow of time in non-Hermitian systems, and apply it to the model in Eq.~(\ref{eq:H_eff}).

Any non-Hermitian Hamiltonian $H_{eff}$ admits a set of right and left eigenstates $H_{eff} | R_j \rangle = E_j | R_j \rangle$,  $\langle L_j | H_{eff} = \langle L_j | E_j^*$, satisfying biorthogonality  $\langle L_j | R_k \rangle = \delta_{jk}$ and resolution of the identity $\sum_j |R_j\rangle \langle L_j| = \mathbb{I}$ (away from exceptional points). Expanding the initial state as
$|\psi_0\rangle = \sum_j |R_j\rangle \langle L_j | \psi_0\rangle$, 
the time-evolved (unnormalized) state reads
\be
|\psi(t)\rangle = \sum_j e^{-i E_j t} |R_j\rangle \langle L_j | \psi_0\rangle.
\ee

Writing the eigenvalues as $E_j = \epsilon_j - i \Gamma_j$, their imaginary part $\Gamma_j$ represents the measurement-induced decay rate of eigenstate $j$. At long times, the dynamics are dominated by the eigenstate(s) with the smallest $\Gamma_j$, which define  the dark, or Zeno, subspace.

The forward probability is 
\be
\begin{split}
P_F &= | | \psi(t)\rangle |^2 = |\sum_j e^{-i E_j t} |R_j \rangle \langle L_j | \psi_0 \rangle|^2 \nonumber \\
&= \sum_k e^{i (E_k^* - E_j)t}  \langle \psi_0|L_k\rangle \langle R_k | R_j \rangle \langle L_j|\psi_0\rangle.
\end{split}
\ee
At long times
\be
P_{F}  \xrightarrow{t \to \infty} |\langle \psi|L_{k_{dark}}\rangle|^2 \langle R_{k_{dark}} | R_{k_{dark}} \rangle e^{- 2 \Gamma_{min} t},
\ee
where $\Gamma_{min}=\Gamma_{k_{dark}} = \min_j \Gamma_j$.

Now consider the backward probability starting from the final state. The final state can be written in terms of the  forward evolution operator  $U=e^{-i H_{eff} t}$ as \be
| \psi(t)\rangle=\frac{U | \psi(t=0)\rangle}{|U | \psi(t=0)\rangle|},
\ee
where the denominator equals $\sqrt{P_F}$. Specializing to the particular form of the nonhermitian Hamiltonian in Eq.~(\ref{eq:H_eff}), and assuming a general time reversal symmetric Hamiltonian generating the unitary dynamics (as the Ising model in Eq.~(\ref{eq:H_eff})),  the backward evolution is generated by the time reversed Hamiltonian. 
Using $\hat{n}=\frac{1-\sigma^z}{2}$ where $\sigma^z$ is odd under time reversal, the time reversed Hamiltonian is
\be
\tilde{H}_{eff} = H_0+ i \frac{\gamma}{2} \sum_{i=1}^L (1-n_i) = H_{eff}+ \frac{i \gamma L}{2}.
\ee
From that we can get the backward propagator 
\[\tilde{U}(0,t) = \theta e^{- i H_{eff}t}\theta^{- 1} = e^{i{\tilde{H}}_{eff}t} = \ e^{i H_{eff}t}e^{- \frac{\gamma}{2}Lt}.\]

Therefore the backward probability is 
\be
P_B=\frac{1}{P_F}   ||   \tilde{U} U | \psi(t=0)\rangle ||^2 = \frac{e^{-L \gamma t}}{P_F}.
\ee 
Thus the arrow of time is 
\be
Q(t) = \log \frac{P_F}{P_B}= \log P_F^2 e^{L \gamma t}.
\ee
At long times 
\be
\label{eq:Sigma_Gamma}
Q(t) \xrightarrow{t \to \infty}  (\gamma L -4 \Gamma_{min})t + const.
\ee
Thus, the steady-state arrow of time in a non-Hermitian dynamics is determined by the imaginary part of the many-body eigenvalue with the smallest decay rate. In particular, $\Gamma_{\min}$ depends on the measurement strength $\gamma$ and vanishes in the limit $\gamma \to 0$, where the evolution becomes unitary. We note that $\Gamma_{\min}$ should not be regarded as a subleading contribution. In fact, as follows from its exact calculation using the Jordan Wigner transformation, it scales extensively with the system size $L$.
\section{Average AoT for continuous measurements}
\label{appendix:mean}
In this appendix we derive Eq.~(\ref{eq:mean_z2}) for the average AoT of a many qubit system under continuous measurements. 

We start from Eq.~(\ref{eqLQ_m_j_tau}),
\be
Q_m=\log \frac{p_m}{\tilde{p}_m} = 2 \epsilon \sum_{\tau=1}^N j_\tau \frac{z_{\tau}+z_{\tau+1}}{2},
\ee 
describing the AoT for a particular single qubit trajectory $m=\{ j_\tau\}$.
While the measurement outcomes $j_\tau$ are biased and depend on the system's state, it is convenient to define
$j_\tau = \epsilon z_\tau + \xi_\tau$, where $\xi$ is an unbiased white noise satisfying 
\be
\label{eq:xi}
\overline{ \xi_\tau } =0, \quad \overline{ \xi_\tau \xi_{\tau'} }=\delta_{\tau \tau'}.
\ee
One thus obtains the desired mean value $\overline{ j_\tau }=p_{+1}-p_{-1}=\epsilon z_\tau$ and variance, up to a correction of order $\epsilon^2$ which vanishes in the continuous measurement limit.

In terms of the $\xi$'s the AoT 
becomes 
\be
\label{eq:Q_result}
Q=2 \epsilon^2 \sum_{\tau=1}^N \left( z_\tau+\frac{\xi_\tau}{\epsilon} \right) \frac{z_\tau+z_{\tau+1}}{2}.
\ee
We define $z_\tau^{\rm{Strat}}=\frac{z_\tau+z_{\tau+1}}{2}$.
In the continuous time limit (with small $\epsilon$ given as above by $\epsilon^2=\gamma dt$) one recovers the result of Dressel et al.~\cite{Dressel_2017},
\be
Q=2 \gamma \int_0^T dt r(t) z(t), ~~~T=N dt,
\ee
with $r(t)=z(t)+\gamma^{-1/2} \tilde{\xi}(t)$ and $\tilde{\xi}(t)=\frac{\xi_\tau}{\sqrt{dt}}$ being the continuous time version of the white Gaussian noise, satisfying $\overline{  \tilde{\xi}(t) \tilde{\xi}(t') }=\delta(t-t')$. 

We now move to $L$ qubits. The expression for the AoT remains intact up to a sum over the contribution from each site $x$ and time step $\tau \to \tau+1$, with $z_{x,\tau}$ being the quantum expectation value of $\sigma^z_{x}$ for the many-qubit wave function at time $\tau$,
\be
Q=2 \epsilon^2 \sum_{x=1}^L\sum_{\tau=1}^N \left( z_{x,\tau}+\frac{\xi_{x,\tau}}{\epsilon} \right) \frac{z_{x,\tau}+z_{x,\tau+1}}{2}.
\ee
Let us evaluate this expression to leading order in $\epsilon$. From causality,  $z_{x,\tau}$ is correlated only with $\xi_{\tau'}$ for $\tau'<\tau$. Therefore we can write
\bea
\label{eq:mean_z3}
\overline{Q}  &=& \sum_{x,\tau} \overline{[2\epsilon^2  z^2_{x,\tau} +\epsilon \xi_{x,\tau} (z_{x,\tau+1}-z_{x,\tau})]}.
\eea
In the second term the contribution $\xi_{x,\tau} z_{x,\tau}$ vanishes by causality. But this differential form is convenient because using the SSE one obtains the increment of the $\sigma^z$ expectation value due to a single measurement step to be
\be
\label{eq:dz}
z_{x,\tau+1}-z_{x,\tau}=\xi_{x,\tau} \epsilon (1-z_{x,\tau}^2).
\ee
Then Eq.~(\ref{eq:mean_z3}) gives
\bea
\overline{Q} = \epsilon^2 \sum_{x,\tau}  \left(  \overline{z_{x,\tau}^2}  + 1\right),
\eea
which is  Eq.~(\ref{eq:mean_z2}). 

\section{Variance of the AoT for continuous measurements
}

\label{appendix:varuance}

  The variance of the AoT is given by $\overline{Q^2} - \overline{Q}^2$ with
\bea
\label{eq:4_terms}
\overline{Q^2} &=& (2 \epsilon^2)^2 \sum_{x,x'=1}^L\sum_{\tau,\tau'=1}^N \times    \\
&& \overline{\left( z_{x,\tau}+\frac{\xi_{x,\tau}}{\epsilon} \right)   z_{x,\tau}^{\rm{Strat}}   \left( z_{x',\tau'}+\frac{\xi_{x',\tau'}}{\epsilon} \right) z_{x',\tau'}^{\rm{Strat}} }. \nonumber
\eea
One obvious term contributing to the variance originates from the equal time and equal space correlation of noise $\langle \xi_{x,\tau} \xi_{x',\tau'} \rangle = \delta_{xx'} \delta_{\tau \tau'}$. This gives the contribution to the variance $(\overline{Q^2} - \overline{Q}^2)_{\langle \xi \xi \rangle}= 4 \epsilon^2 \sum_x \sum_\tau \overline{z_{x,\tau} z_{x',\tau'}}$.   This contribution to the variance was applied as an approximation for a single qubit. In the time-continuous formulation it reads ${\rm{var}}(Q) |_{\langle \xi \xi \rangle}=4\gamma \int_0^T dt \overline{z(t)^2}$.

However, in the many-body case having ``entangling" phases, it is important to keep terms describing long range correlations. From Eq.~(\ref{eq:4_terms}) we  have
\be
\label{eq:4_terms1}
{\rm{var}}(Q)={\rm{var}}(Q) |_{\langle z z \rangle}+2{\rm{var}}(Q) |_{\langle \xi z \rangle} 
+{\rm{var}}(Q) |_{\langle \xi \xi \rangle},
\ee
where 
\bea
{\rm{var}}(Q) |_{\langle z z \rangle} &=&
(2 \epsilon^2)^2 \sum_{x,x'=1}^L\sum_{\tau,\tau'=1}^N \overline{ z_{x,\tau}   z_{x,\tau}    z_{x',\tau'} z_{x',\tau'} }, \nonumber \\ 
{\rm{var}}(Q) |_{\langle \xi z \rangle} &=&
4\epsilon^3 \sum_{x,x'=1}^L\sum_{\tau,\tau'=1}^N \overline{ \xi_{x,\tau}   z_{x,\tau}    z_{x',\tau'} z_{x',\tau'} }.
\eea

However, for the variance, in addition to the local term denoted ${\rm{var}}(Q) |_{\langle \xi \xi \rangle}$ in Eq.~(\ref{eq:4_terms1}), there are the other terms which are sensitive to the long distance behavior  of this correlator.

\section{Derivation of Eq.~(\ref{eq:w_d}) for $w^{(n)}_{d}(\sigma,\tau)$}
\label{appendix:w_d}
\begin{figure}
\centering
\includegraphics[width=\columnwidth]{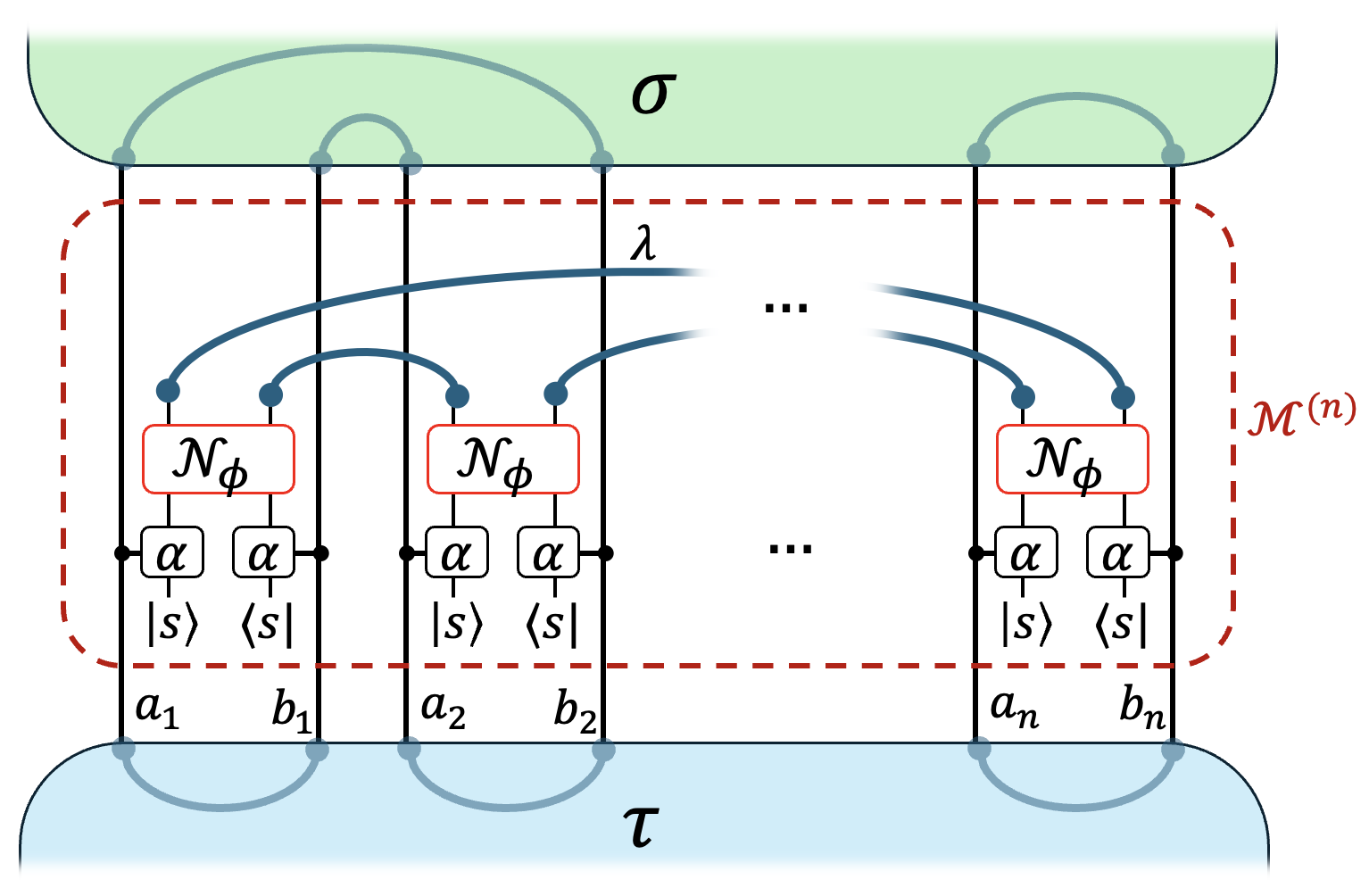}
\caption{Tensor network representation of the weight $w_d^{(n)}(\sigma,\tau)$. Here $\sigma,\tau$ are permutations of $[n]$.
The relative permutation of the example $\sigma$ and $\tau$ in the figure is shown to have at least 2 cycles where one is $a_1=b_1=a_2=b_2$ and the other is $a_n=b_n$. The index $\lambda$ is the trace index over the ancilla degrees of freedom. It is a single summation since the dephased density matrix of each ancilla is diagonal.}
\label{fig_w_d}
\end{figure}

We closely adopt the calculation in appendix A of Ref.~\cite{Bao_2020} to our measurement operator. The definition of the weights $w_d^{(n)}(\sigma,\tau )$ is described by the tensor network shown in Fig.~\ref{fig_w_d} and can be written as the trace of the diagonal link for a given pair of permutations $\sigma$ and $\tau$,
\be
\label{eq:w_d_trace_def}
w_d^{(n)}(\sigma, \tau) = \sum_{\boldsymbol{ab} }\hat{\tau}_{\boldsymbol{ab}}\hat{\sigma}_{\boldsymbol{cd}} \mathcal{M}^{(n)}_{\boldsymbol{{ab;cd}}},
\ee
with the bold symbols being vectors of $n$ indices for summation over $q$ dimensions, e.g. $\boldsymbol{a} = (a_1,a_2,\cdots,a_n),~\boldsymbol{b} = (b_1,b_2,\cdots,b_n)$, and the tensors $\hat{\tau}$ and $\hat{\sigma}$ enforce the permutations $\tau$ and $\sigma$ between their 2 vectors of indices. From Fig.~\ref{fig_w_d} it is immediately clear that $\mathcal{M}^{(n)}_{\boldsymbol{{ab;cd}}}$ is diagonal in the computational basis, that is, it does not mix the indices between $\hat{\tau}$ and $\hat{\sigma}$ and thus could be simplified to 
\be
\label{eq:w_d_trace_def}
w_d^{(n)}(\sigma, \tau) = \sum_{\boldsymbol{ab} }\hat{\tau}_{\boldsymbol{ab}}\hat{\sigma}_{\boldsymbol{ab}} \mathcal{M}^{(n)}_{\boldsymbol{{ab}}}.
\ee
The explicit form of $\mathcal{M}^{(n)}_{\boldsymbol{{ab}}}$ is given by tracing over all ancilla degrees of freedom, after they are entangled with the system qudits by $\hat{R}_\alpha$ and goes through the dephasing gate $\mathcal{N}_\phi$, as seen in Fig.~\ref{fig_w_d}. The dephased density matrix $\rho^{(k)}_{a_k b_k}$ of a single ancilla on the $k$-th replica is thus the following diagonal matrix

\be
\mathcal{N}_\phi[\rho^{(k)}_{a_k b_k}] = \sum_{\lambda=1}^q  |\lambda \rangle \langle  \lambda| e^{-i\hat{Y}_{a_k} \alpha} |s\rangle \langle s | e^{i\hat{Y}_{b_k}\alpha}|\lambda \rangle \langle  \lambda| .
\ee
Those matrix elements were already calculated in Eq.~(\ref{eq:matrix_element}) and give $\langle \lambda| e^{-i\hat{Y}_{i} \alpha} |s\rangle = c_0 + c_1 \delta_{\lambda, i}$. Tracing over the product of the $n$ copies of those matrices gives

\be
\mathcal{M}^{(n)}_{\boldsymbol{{ab}}} = \sum^q_{\lambda=1} \prod^n_{k=1} (c_0 + c_1 \delta_{\lambda, a_k})(c_0 + c_1 \delta_{\lambda, b_k}),
\ee
since we can change the order of terms in the product, for any given permutation $\sigma$ we can order the right term by $b_{\sigma(k)}$ instead of $b_k$.
Then we notice from Eq.~(\ref{eq:w_d_trace_def}) that the $\hat{\sigma}_{\boldsymbol{ab}}$ tensor exactly enforces $b_{\sigma(k)} = a_k$ in the summation so both terms in the product are equal and reduce to their square giving the weight of
\be
w_d^{(n)}(\sigma, \tau) = \sum_{\boldsymbol{ab} }\hat{\tau}_{\boldsymbol{ab}}\hat{\sigma}_{\boldsymbol{ab}} \sum^q_{\lambda=1} \prod^n_{k=1} (c_0^2 + (c_1^2 + 2 c_0 c_1) \delta_{\lambda, a_k}).
\ee

We proceed with noticing that now the summand has no dependence in the $\boldsymbol{b}$ indices other than in the permutation tensors, we can thus change the order of summation over $\boldsymbol{b}$ to summation over $\tau^{-1}(\boldsymbol{b})$. The $\hat{\tau}$ tensor then restricts the summation to $\boldsymbol{a} = \tau(\tau^{-1}(\boldsymbol{b})) = \boldsymbol{b}$ and the $\hat{\sigma}$ tensor restricts the summation to $\boldsymbol{a} = \sigma(\tau^{-1}(\boldsymbol{b})) = \xi (\boldsymbol{b})$, with $\xi=\sigma\cdot\tau^{-1}$ the relative permutation. Resulting in a summation only over $\boldsymbol{a}$, restricted to $\boldsymbol{a} = \xi(\boldsymbol{a})$, that is, restricted to the cycles of $\xi$. 

Suppose that the relative permutation $\xi$ has $k(\xi)$ cycles (denoted $k$) of lengths $\{ n_1, n_2,\dots, n_k\}$ such that $\sum_{i=1}^k n_i=n$. Then the restricted sum $\sum_{\boldsymbol{a}=\xi(\boldsymbol{a})}$ could be  written as $k$ independent sums over $j_1, j_2, \dots, j_k$ each corresponding to a separate cycle. Each $j_\ell$ stands for $n_\ell$ indices of $\boldsymbol{a}$ that are equal due to $\boldsymbol{a}=\xi(\boldsymbol{a})$ so the weight simplifies to 
\bea
w_d^{(n)}(\sigma, \tau) = \sum_{\boldsymbol{a}=\xi(\boldsymbol{a}) }\sum^q_{\lambda=1} \prod^n_{k=1} (c_0^2 + (c_1^2 + 2 c_0 c_1) \delta_{\lambda, a_k})
\nonumber
\\
= \sum_{j_1, \cdots, j_k}\sum^q_{\lambda=1} \prod^k_{\ell=1} (c_0^2 + (c_1^2 + 2 c_0 c_1) \delta_{\lambda, j_\ell})^{n_\ell}~~
\nonumber
\\
= \sum_{\lambda=1}^q  \prod_{\ell=1}^k \left[\sum_{j_\ell=1}^q \left( c_0^2+  (c_1^2+2 c_0 c_1) \delta_{\lambda, j_\ell} \right)^{n_\ell} \right], ~~
\eea
with the last transition coming from the dependence of each term in the product in only one of the $j_\ell$ indices. The sum over $j_\ell$ is solved by noticing that for each $\lambda$ there are $q-1$ terms for which $\delta_{\lambda, j_\ell}=0$ and a single term that is resolved to $\left( c_0+c_1  \right)^{2n_\ell}$. The sum over $\lambda$ could then be trivially solved to a factor of $q$ to reach the final expression

\be
w_d^{(n)}(\sigma,\tau )= q  \prod_{\ell=1}^k \left[ \left( c_0 + c_1  \right)^{2n_\ell} +(q-1)  \left( c_0 \right)^{2n_\ell}\right], 
\ee
which is Eq.~(\ref{eq:w_d}) upon substituting $c_0=\frac{\cos \alpha }{\sqrt{q}}
-\frac{\sin \alpha }{\sqrt{q(q-1)}}$ and $c_1=
\sin \alpha \sqrt{\frac{q}{q-1}}$.

\section{Generalized mapping to percolation}
\label{appendix:FK}
\begin{figure}
\centering
\includegraphics[width=\columnwidth]{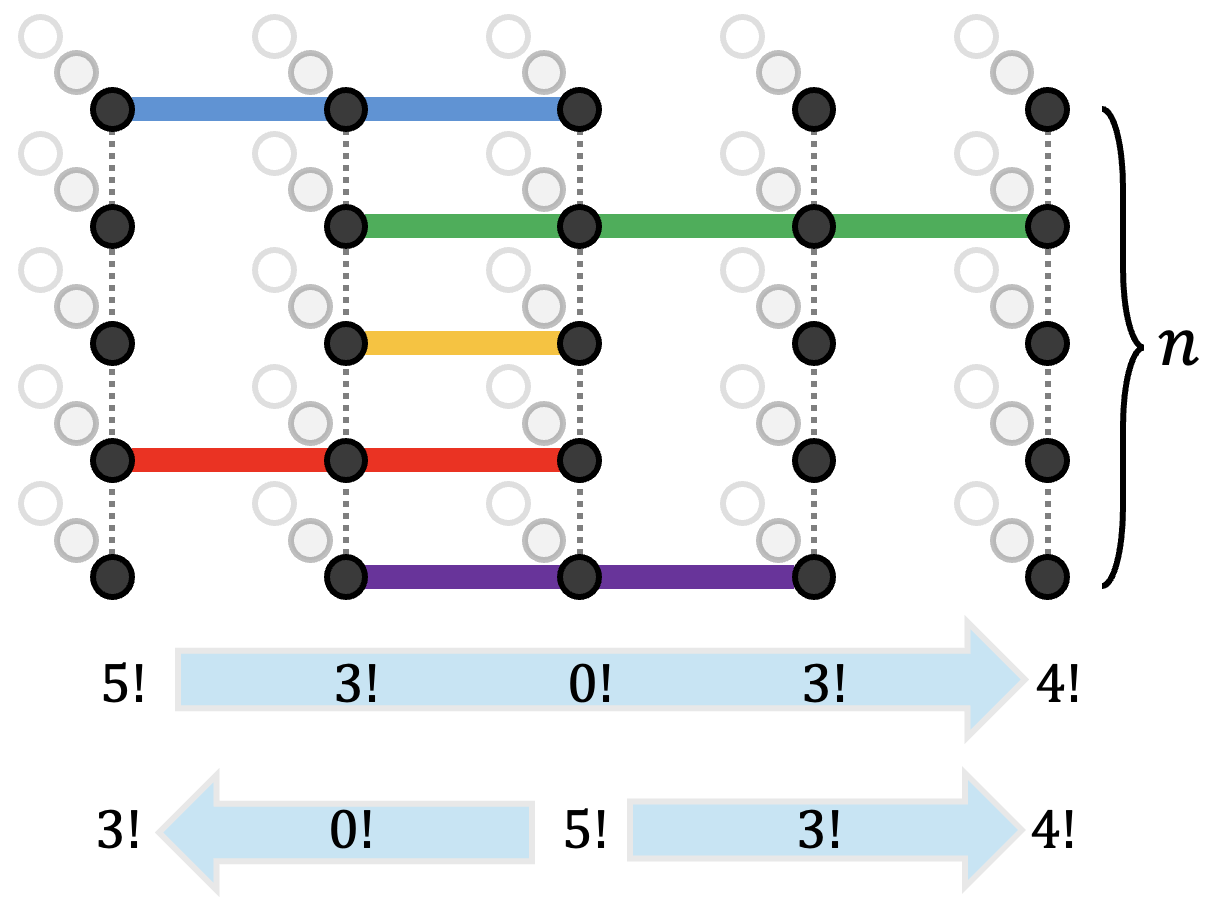}
\caption{Graphical representation of Eq.~(\ref{eq:b}) for $n=5$. For each bond $\langle xy \rangle$, the value of $b_{xy}^{(i)}$ sets whether or not an edge exists in the $i$-th floor corresponding to the $i$-th color. By overlapping the different colors, we see a single cluster (1-dimensional for simplicity). In both orders of counting we find the total number of valid permutations to be $5!\cdot4!\cdot(3!)^2$ or in the general form $(5!)^1(4!)^1(3!)^2(2!)^0(1!)^0(0!)^1$.}
\label{fig:colored_clusters}
\end{figure}

Let us now analyze the model defined by $\tilde{Z}^{(n)}$ in Eqs.~(\ref{eq:Ztilde}) and (\ref{eq:Weight}), 
\be
\tilde{Z}^{(n)}=\sum_{\{ \sigma \}} \prod_{\langle x y \rangle}
(\sin^2 \alpha)^{\,n-n_{{\rm{fp}}}(\sigma_x \sigma_y^{-1})} \, ,
\ee
where $n_{\rm{fp}}(\sigma_x \sigma_y^{-1}) = \# \{ i \; | \; \sigma_x(i)=\sigma_y(i) \}$. 
We consider a square (or diamond) lattice with $NL$ edges. We can rewrite the bond weight as
$W(\sigma_x,\sigma_y)=\frac{(1+v)^{n_{\rm{fp}}(\sigma_x \sigma_y^{-1})}}{(1+v)^{n}}$ with $1+v=\sin^{-2}\alpha$.
Using
\[
n_{\rm{fp}}=\sum_{i=1}^n \delta_{\sigma_x(i)\sigma_y(i)},
\]
the partition function can be written as
\be
\tilde{Z}^{(n)}=\frac{1}{(1+v)^{NLn}} \sum_{\{\sigma\}}
\prod_{(\langle xy\rangle,i)}
\left(1+v\,\delta_{\sigma_x(i)\sigma_y(i)}\right),
\ee
where $(\langle xy\rangle,i)$ denotes an edge--color pair. Thus, each edge--color pair contributes a corresponding factor. We now expand each factor by introducing a bond variable $b_{xy}^{(i)}=0,1$,
\be
1+v\,\delta_{\sigma_x(i)\sigma_y(i)}
=\sum_{b_{xy}^{(i)}=0}^1 \left(v\,\delta_{\sigma_x(i)\sigma_y(i)}\right)^{b_{xy}^{(i)}}.
\ee
The partition function then becomes
\be
\label{eq:b}
\tilde{Z}^{(n)}= \frac{1}{(1+v)^{NLn}}
\sum_{\{ b \}} v^{\sum b_{xy}^{(i)}}
\sum_{\{ \sigma \}}
\prod_{(\langle x y\rangle, i):\, b_{xy}^{(i)}=1}
\delta_{\sigma_x(i)\sigma_y(i)} .
\ee
The delta-function constraints imply that for each $i$, edges with
$b_{xy}^{(i)}=1$ enforce $\sigma_x(i)=\sigma_y(i)$. This is illustrate in Fig.~\ref{fig:colored_clusters} where $i=1,\dots,n$ corresponds to floors which also have different colors. The bonds for which $b_{xy}^{(i)}=1$ are the existing edges and those with $b_{xy}^{(i)}=0$ are absent.  In fact, the sum over ${ b }$ is equivalent to a sum over all colored graphs,
\be
\sum_{ \{ b \}} = \sum_{{ G }},
\ee
where $G$ consists of $n$ subgraphs, one for each color $i$, and the occupied bonds in the subgraph of color $i$ are those bonds for which $b_{xy}^{(i)}=1$. We can also denote the term
\be
\mathcal{V}(G) = \sum_{\{ \sigma \}}
\prod_{(\langle x y\rangle, i):\, b_{xy}^{(i)}=1}
\delta_{\sigma_x(i)\sigma_y(i)} ,
\ee
which, for a given graph $G$, counts the number of valid permutation assignments that satisfy the coloring constraints. 

The rest of the terms reduce to the probability of the graph $G$. Let us denote the total number of occupied bonds in the colored graph by $|E|=\sum_{\langle xy\rangle,i} b_{xy}^{(i)}$. 
The maximal possible number of bonds is $N L n$. 
We define the bond occupation probability as $p = \frac{v}{1+v} = \cos^2 \alpha$. 
Thus the $v$-dependent terms become the probability of each graph which is simply $P(G)=p^{|E|} (1-p)^{N L n - |E|}$ and lead to the simple form 
\be
\tilde{Z}^{(n)} = \sum_G P(G) \mathcal{V}(G).
\ee
Noticing that $\mathcal{V}(G)$ goes to $1$ in the limit of $n\to1$, we obtained a partition function that seems to describe percolation in the replica limit $n\to1$ (a colorless-lattice). Since the colors are independent, the probability that no color occupies an edge $\langle xy \rangle$ is 
\[
p(\text{no color on }\langle xy \rangle) = (1-p)^n,
\] 
so that the probability that at least one color occupies the edge is 
\[
p(\text{at least one color on }\langle xy \rangle) = 1-(1-p)^n.
\] 
Therefore, in this limit the projected graph reduces to ordinary bond percolation on the original lattice, with effective bond probability $p_{\rm eff} = 1-(1-p)^n$. Taking the analytic limit $n \to 1$ gives $p_{\rm eff} = p + \mathcal{O}(n-1)$, showing that $P(G)$ has a vanishing dependence on $n$ in the replica limit. The resulting partition function satisfies $\tilde{Z}^{(1)} = 1$, since it is simply the sum over probabilities.

We  now analyze $\mathcal{V}(G)$, the number of valid permutations on a graph $G$. An intuitive approach is to go over each cluster in the colored-graph, with a cluster being a set of vertices connected by at least one of the colors. See Fig.~\ref{fig:colored_clusters} for an example of a cluster of 5 vertices in a 5-color graph.

Since each cluster is completely unconstrained from the rest and has at least one vertex, it follows that it has at least $n!$ valid permutation assignments. To count the  possible assignments per cluster we can simply iterate over all vertices in the cluster. The first vertex is unconstrained and contributes the factor of $n!$, but any of its neighbors now add a factor of $\ell!$ with $0\le\ell\le n-1$. Notice that the order of iteration may change the factor any given vertex is contributing, but the overall count is of course independent of the chosen order (though it must span from the first vertex, going from neighbor to neighbor). 

The number of valid assignments for cluster $i$ must then be $\prod_{\ell=0}^{n} (\ell!)^{k_{i,\ell}}$ with $\sum_{\ell=0}^{n}k_{i,\ell} = |C_i|$ and $|C_i|$ the size of the $i$-th cluster. Most importantly, for all clusters $k_{i,n} = 1$, meaning that one and only one vertex per cluster contributes the full $n!$ factor.
In Fig.~\ref{fig:colored_clusters} we show an example of counting the number of valid permutations in a single cluster, in 2 different ways.

Counting those factors over all clusters we get
\be
\mathcal{V}(G) = \prod_{\ell=0}^{n} (\ell!)^{K_{\ell}},
\ee
with $K_\ell = \sum_{\text{cluster-}i} k_{i,\ell}$, that satisfies $\sum_{\ell=0}^n K_\ell = NL$ This means that $K_n \equiv \#\mathcal C_G$ is the number of clusters in the graph $G$, and thus for $n\to1$ we can already see that $K_{n-1} \to NL-\#\mathcal C_G$ which will be useful soon.

The derivative with respect to $n$, which gives the arrow of time, is $\partial_n \log \tilde{Z}^{(n)} 
= \frac{1}{\tilde{Z}^{(n)}}\partial_n \tilde{Z}^{(n)}$. Since $\tilde{Z}^{(1)}=1$, we get
\be
\label{eq:partition_color}
\lim_{n\to 1}\partial_n \log \tilde{Z}^{(n)}  = \lim_{n\to 1} \sum_{G} P(G) \partial_n \mathcal{V}(G).
\ee
Focusing on $\partial_n \mathcal{V}(G)$ we can write:
\bea
\partial_n \mathcal{V}(G) &=& \mathcal{V}(G) \partial_n \log(\mathcal{V}(G)) \nonumber
\nonumber
\\ 
&=& \mathcal{V}(G) \partial_n (K_{n}\log(n!) + K_{n-1}\log((n-1)!) + \cdots )
\nonumber
\\
&=& \mathcal{V}(G) \Bigl( K_{n}\frac{\partial_n (n)!}{(n!)} +  K_{n-1}\frac{\partial_n (n-1)!}{(n-1)!} + \cdots
\nonumber
\\
&+&   (\partial_n K_{n}) \log(n!) +  (\partial_n  K_{n-1}) \log((n-1)! + \cdots\Bigr).
\nonumber
\\
\eea
In the replica limit we can already tell that only the $n$ and $n-1$ terms will survive. Looking at the first kind of terms in the replica limit we get
\bea
\lim_{n\to1}\left(K_{n}\frac{\partial_n (n)!}{(n!)} +  K_{n-1}\frac{\partial_n (n-1)!}{(n-1)!}\right) \nonumber
\\
= \#\mathcal C_G (1-\gamma) + (NL-\#\mathcal C_G) (-\gamma)\nonumber
\\
=
\#\mathcal C_G -\gamma NL,
\eea
with $\gamma$ the Euler–Mascheroni constant.
The other type of terms vanish if $ \partial_n K_{n}$ and $ \partial_n K_{n-1}$ both resolve to constants. Expanding near $n=1$ we have
\bea
 (\partial_n K_{n}) \log(n!) +  (\partial_n K_{n-1}) \log((n-1)!)
\nonumber
\\ 
= \left[ (\partial_n K_{n}) (1-\gamma) +  (\partial_n K_{n-1})(-\gamma)\right](n-1).
\eea
So this could be resolved to a constant if the analytical continuation of either $\partial_n K_{n}$ or $\partial_n  K_{n-1}$ diverge as $\sim(n-1)^{-1}$. Let us denote this possible term as $R$ and return to it. 
Going finally back to the arrow of time in Eq.~(\ref{eq:partial_n}) and using Eq.~(\ref{eq:partition_color}) we reach

\be
\langle\langle\overline{Q}\rangle\rangle = 2\langle \#\mathcal C_G \rangle -(2\gamma + \log(p))NL + 2\langle R \rangle.
\ee
Here $\langle\cdot\rangle$ denotes average over graphs. So the first term is twice the average number of clusters. Crucially, this term already gives rise to the nonanalytic behavior with exponent $2-\alpha$ of the AoT.
Since the arrow of time must vanish for $\alpha=0;~p=1$, and in this limit $\langle \#\mathcal C_G \rangle=\mathcal{O}(1)$,
\be
\langle\langle\overline{Q}\rangle\rangle = 0 = - 2\gamma NL + 2\langle R \rangle + \mathcal{O}(1).
\ee
We must infer that in the weak measurement limit $\langle R \rangle \approx \gamma NL$, which can be achieved simply if $\lim_{n\to1} (\partial_nK_{n-1}) = -LN (n-1)^{-1}$. Whether or not other terms can appear through this correction is unclear. Though we emphasize that only terms that diverge as $(n-1)^{-1}$ in $ (\partial_n K_n);~ (\partial_n K_{n-1})$ will contribute. Constants or weaker divergence terms vanish and stronger divergence term should not exists since the AoT must be finite for non-projective measurements. We thus reach the final form of the AoT,

\be
\langle\langle\overline{Q}\rangle\rangle = 2\langle \#\mathcal C_G \rangle - NL \log p + \dots~.
\ee
Most importantly, it contains a term that has non-analytic behavior over the MIPT (see main text at subsection \ref{subsec:percolation}).

In Fig.~\ref{fig:n_p} we compute $n_c$ via Monte Carlo sampling on small graphs. The simulated graphs $G=(V,E)$, including boundary conditions, consist of 3 subgraphs (see lower insets): a bulk subgraph and two boundary subgraphs. The bulk subgraph is the diamond lattice with periodic boundary conditions along the horizontal direction. It contains $N-1$ rows, each having $L/2$ vertices. The top boundary subgraph is a row of $L$ vertices (see small dots). Each of the $L/2$ vertices in the top row of the bulk subgraph are attached with edges to 2 vertices of the top boundary subgraph. The bottom boundary subgraph is defined in the same way. The Monte Carlo sampling creates subgraphs of $G$ where bonds are present with probability $p$. A remark on boundary conditions should be made: The permutations in the upper boundary are fixed to the identity, while they are completely free at the bottom boundary~\cite{Bao_2020}. Therefore, the clusters that actually contribute to $n_c$ are those clusters which do not have any vertex in common with the top boundary. 

\section{Distribution of $\langle\hat{P}_j\rangle$ for a Haar-random state}

We consider a random state $|\psi\rangle$ of $d$ dimensions, that is not normalized, where each basis component $c_i$ is generated by a complex gaussian distribution $\mathcal{CN}$ with zero mean,
\be
|\psi\rangle = \sum_{i=1}^d c_i |i\rangle~~,~~c_i\sim\mathcal{CN}(0,\sigma).
\ee
The standard deviation $\sigma$ will turn out to be unimportant so we will set $\sigma=1$ for simplicity.
Such a state is Haar-random due to the invariance of the joint  distribution to any unitary transformation, $~p(c_1, c_2, \cdots) \propto e^{-\frac{1}{2}\langle\psi|\psi\rangle}$. Normalization maintains the invariance as well, so that the eventual state is indeed Haar-random.

The squared absolute values of the components are then Gamma-distributed. This is shown by writing explicitly $x_i = \Re\{c_i\}\sim \mathcal{N}(0,\frac{1}{2}), ~~ y_i =\Im\{c_i\}\sim \mathcal{N}(0,\frac{1}{2})$
then changing variables to $X_i = x_i^2,~Y_i=y_i^2$ the new probability distribution function is
\be
p(X_i) = 2\cdot p(x_i)\left|\frac{dx_i}{dX_i}\right| 
= \sqrt{\frac{2}{\pi}}X_i^{-\frac{1}{2}} e^{2X_i},
\ee
with the factor of 2 coming from considering both negative and positive values of $x_i$. This is exactly a gamma distribution with $\alpha=1/2, \beta=2$. In other words $X_i, Y_i\sim {\rm{Gamma}}(\frac{1}{2},2)$.

The expectation value of any projection operator $\langle\hat{P}\rangle$ is the sum of a subset of squared components, together with the  normalization factor. Without loss of generality we will assume this subset is the first $n$ squared components ($n<d$),
\be
\langle\hat{P}\rangle = \frac{\sum_{i=1}^n |c_i|^2}{\sum_{i=1}^d |c_i|^2} = \frac{\sum_{i=1}^n |c_i|^2}{\sum_{i=1}^n |c_i|^2 + \sum_{i=1}^{d-n} |c_{i+n}|^2} \equiv \frac{A}{A+B}.
\ee
We thus have 2 random variables, $A$ and $B$, that are a sum of $2n$ and $2(d-n)$ ${\rm{Gamma}}$-distributed values. Those are also ${\rm{Gamma}}$ distributed with the new $\alpha$ parameter being the sum of $\alpha$'s, $A\sim {\rm{Gamma}}(n, 2), ~B\sim {\rm{Gamma}}(d-n, 2)$. Then using a known relation between the quotient of Gamma distributions of equal $\beta$ parameter, to the Beta distribution, we get $\langle P_j \rangle \sim \text{Beta}(n, d-n)$.

For $L'$ qudits of dimension $q$ we get $d=q^{L'}$ and any projector $\hat{P}_j$ of a single-qudit state will have $n=d/q = q^{L'-1}$. We finally find that for such a Haar-random state, the distribution of expectation values of any local projector is $\langle P_j \rangle \sim {\rm{Beta}}(q^{L'-1}, q^{L'}-q^{L'-1})$. With the variance of such distribution being 
\be
\text{Var}[{\langle P_j \rangle}] = \frac{1-\frac{1}{q}}{q^{L'+1}+q} \xrightarrow[q \to \infty]{} \frac{1}{q^{L'+1}}.
\ee

\end{appendix}


\bibliography{bibliography}


\end{document}